\documentclass[a4paper,11pt]{article}
\pdfoutput=1

\usepackage{jheppub}

\usepackage{subfig}
\usepackage{xspace}
\usepackage[countmax]{subfloat}
\usepackage{slashed}

\usepackage{comment}

\usepackage{bbm}

\newcommand{\as}{\alpha_s}

\def\log{\text{log}}

\def\be{\begin{equation}}
\def\ee{\end{equation}}

\DeclareRobustCommand{\Sec}[1]{Sec.~\ref{#1}}

\newcommand{\tpp}{t_\perp}
\newcommand{\tpa}{t_\parallel}
\newcommand{\rad}{\mathcal{R}}

\usepackage{color}
\definecolor{darkblue}{rgb}{0,0,0.5}
\definecolor{darkred}{rgb}{0.5,0,0}
\definecolor{darkgreen}{rgb}{0,0.5,0}

\title{Safe Use of Jet Pull}

\author[1]{Andrew Larkoski,}
\affiliation[1]{Physics Department, Reed College, Portland, OR 97202,USA}

\author[2]{Simone Marzani,}
\affiliation[2]{Dipartimento di Fisica, Universit\`a di Genova and INFN, Sezione di Genova,\\ Via Dodecaneso 33, 16146, Italy}
\author[2]{and Chang Wu}

\emailAdd{larkoski@reed.edu}
\emailAdd{simone.marzani@ge.infn.it}
\emailAdd{chang.wu@ge.infn.it}

\abstract{
Jet pull is an observable designed to probe colour flow between jets. 
Thus far, a particular projection of the pull vector, the pull angle, has been employed to distinguish colour flow between jets produced by a colour singlet or an octet decay. 
This is of particular importance in order to separate the decay of a Higgs boson to a pair of bottom quarks from the QCD background. 
However, the pull angle is not infra-red and collinear (IRC) safe.  
In this paper we introduce IRC safe projections of the pull vector that exhibit good sensitivity to colour flow, while maintaining calculability. We calculate these distributions to next-to-leading logarithmic accuracy, in the context of the hadronic decay of a Higgs boson, and compare these results to Monte Carlo simulations.
This study allows us to define an IRC safe version of the pull angle in terms of asymmetry distributions. 
Furthermore, because of  their sensitivity to wide-angle soft radiation, we anticipate that these asymmetries can play an important role in assessing subleading colour correlations and their modelling in general-purpose Monte Carlo parton showers. 

}

\begin{document} 
\maketitle

\section{Introduction}\label{sec:intro}

During this long shutdown phase, the experiments of the CERN Large Hadron Collider (LHC) are gearing up for the third run of the accelerator. While the increase in centre-of-mass energy will be modest, the path to discovery of new physics, which thus far has proven so elusive, will likely involve careful analyses of large dataset, in order to expose subtle deviations from Standard Model (SM) predictions.
Together with the search for beyond the Standard Model (BSM) particles or interactions, careful studies of the Higgs sector will continue to constitute the second, but equally important, leg of the LHC physics program. In particular, pinning down the couplings of the Higgs boson to the fermions may lead to a deeper understanding of the flavour structure of the SM. 
In this context, both the ATLAS and CMS collaborations have reached the sought-for statistical significance for the decay of the Higgs into bottom quarks~\cite{Aaboud:2018zhk,Sirunyan:2018kst} in Run~II data. 

Typical events from proton-proton collisions at the LHC are filled with strongly-interacting particles, the dynamics of which is described by Quantum Chromo Dynamics (QCD). 
It follows that QCD radiation has a profound impact on both BSM and Higgs physics. The reason is twofold. Firstly, SM processes involving quarks and gluons often constitute the main background, which often dwarves the signal of interest by orders of magnitude. Furthermore, QCD radiation often accompanies the production of the particles of interest, and indeed it offers valuable handles to study them; e.g.\ Higgs production in association with jets. 
In our current study we concentrate on the latter issue, namely we discuss observables that by measuring QCD radiation in a signal event, provide us with information on the properties of the particle we are studying. In particular, we are interested in assessing the colour quantum numbers of a resonance decaying into quarks. This is of clear interest for BSM searches but it also provides a useful handle in distinguishing the decay of a colour singlet (such as the Higgs) into quarks from the decay of a colour octet (such as the gluon) in the same final state.

A powerful observable that is able to probe colour flow is jet pull, which was first proposed in Ref.~\cite{Gallicchio:2010sw}. Since then, a number of experimental analyses has been devoted to this observable: from a pioneering measurement performed by the D$\emptyset$ collaboration at the Tevatron~\cite{Abazov:2011vh}, to two measurements performed by the ATLAS collaboration at the LHC, at centre-of-mass energy of~8~TeV~\cite{Aad:2015lxa} and 13~TeV~\cite{Aaboud:2018ibj}. Most of the measurements concentrate in a particular projection of the jet pull vector, the so-called pull angle, that would, in principle offer the best sensitivity. However, as the experimental uncertainties on the measurement grew smaller, it became apparent that general-purpose Monte Carlo parton showers struggled in modelling the pull angle distribution. In particular, it has been pointed out that the datapoints corresponding to the measurement of the pull angle in $W$ decay are almost equidistant from the result obtained from a standard Monte Carlo simulation and from a simulation where the $W$ is assumed to be a colour octet~\cite{Aaboud:2018ibj}.

In a previous Letter~\cite{Larkoski:2019urm}, we embarked in a detailed study of the pull angle distribution, with the hope that analytic resummation could shed light on those discrepancies. While our perturbative prediction, supplemented with an estimate of a non-perturbative contribution, could describe the experimental data, it still suffered from large theoretical uncertainties, rendering any firm conclusion difficult to draw. The main bottleneck of the theoretical calculation resides on the fact that the pull angle distribution is not infra-red and collinear (IRC) safe but only Sudakov safe~\cite{Larkoski:2013paa,Larkoski:2014wba,Larkoski:2015lea}. Because the theoretical understanding of Sudakov safe observables is still in its infancy, it is not clear how theoretical accuracy can be achieved (and rigorously assessed) beyond the first order. Furthermore, while IRC safety ensures the presence of a kinematical region where non-perturbative effects are genuine power corrections, no such guarantee exists for Sudakov-safe observables and consequently, non-perturbative physics can contribute to the observable as an order-one effect. 
In this paper we overcome these difficulties by defining suitable projections of jet pull that share many of the desirable features of the pull angle, but at the same time are IRC safe. This enables us to perform perturbative calculations at a well-defined, and in principle improvable, accuracy.  

The paper is organised as follows. In Section~\ref{sec:obs} we recall the definition of jet pull and we introduce the safe projections we want to study. Section~\ref{sec:theory} contains the all-order calculations for the observables of interest, while in Section~\ref{sec:pheno} we perform phenomenological studies, which include a comparison to the results obtained using Monte Carlo event generators. In Section~\ref{sec:asym} we exploit the theoretical understanding achieved so far to introduce novel asymmetry observables that aim to better probe colour flow in an infra-red and collinear safe way. Finally, we draw conclusions in Section~\ref{sec:conclusions} and outline our plan for future work on this topic. 

\newpage

\section{Jet Pull}\label{sec:obs}
\begin{figure}
\centering
\includegraphics[width=0.8 \textwidth]{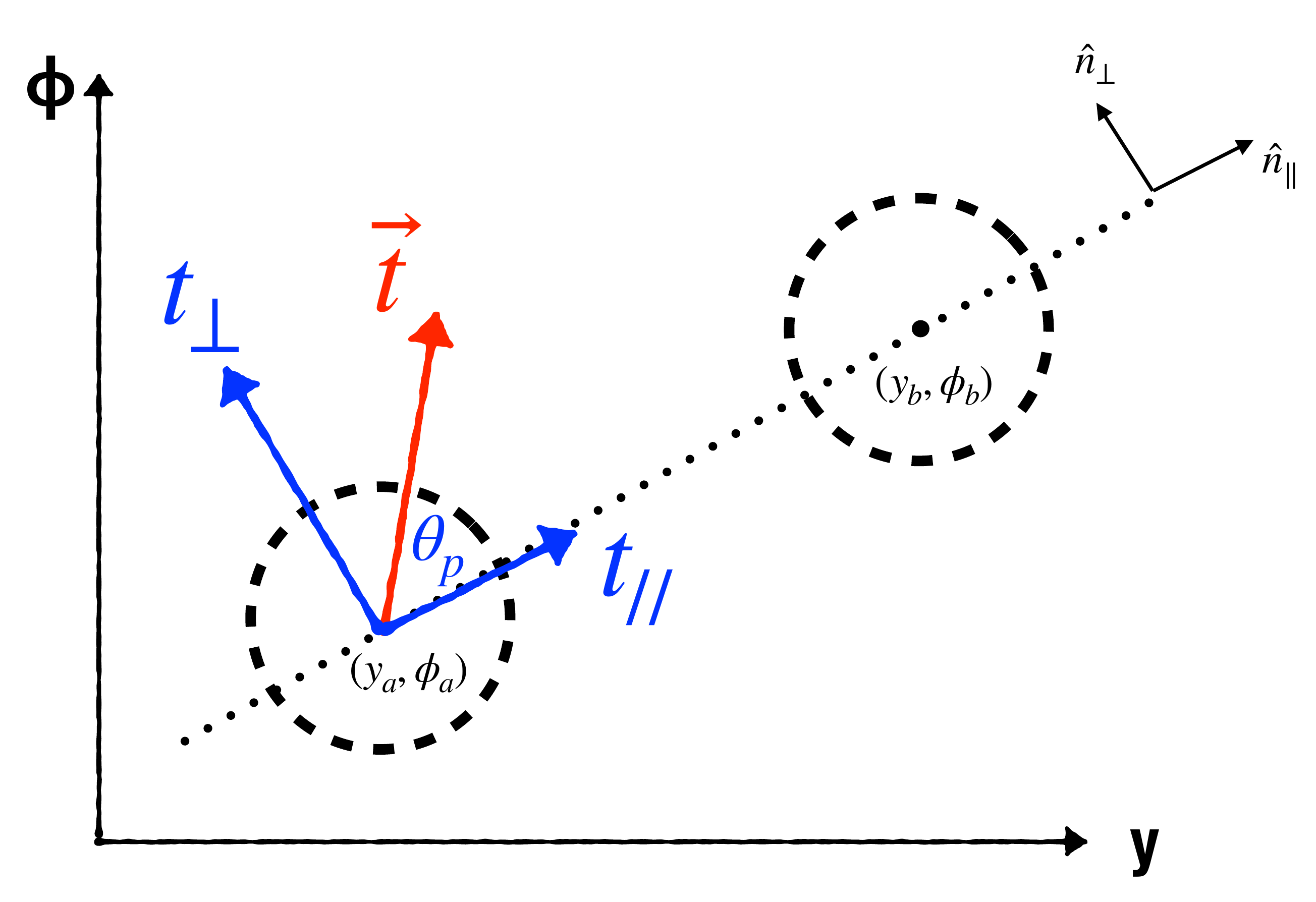}
\caption{A cartoon representation of the rapidity ($y$) and azimuth ($\phi$) plane with the pull vector of jet $a$ and its projections, which are defined with respect the jet $b$.}
\label{rap-azim} 
\end{figure}

The pull vector is a two-dimensional jet shape defined as~\cite{Gallicchio:2010sw}
\begin{equation}\label{pull-def}
\vec t= \frac{1}{p_{ta}}\sum_{i\in J}p_{ti} |\vec{r}_i|^2 \hat r_i\,,
\end{equation}
where the sum runs over all particles in a jet and
\begin{equation}\label{ri-def}
\vec{r}_i=(y_i-y_a, \phi_i-\phi_a), \quad \text{and} \quad \hat r_i= \frac{\vec{r}_i}{|\vec{r}_i|}.
\end{equation}
The coordinates of the jet centre in the rapidity-azimuth plane are $(y_a,\phi_a)$ and $p_{ta}$ is the jet transverse momentum. We are interested in measuring the pull of jet $a$ in the presence of a second jet $b$, that we center at $(y_b,\phi_b)$. To this purpose, we find useful to introduce the two unit vectors
\begin{align}\label{unit-vectors}
\hat{n}_\parallel&= \frac{1}{\sqrt{\Delta y^2 + \Delta \phi^2}} (\Delta y, \Delta \phi)=(\cos \beta, \sin \beta)
, \nonumber \\
\hat{n}_\perp&= \frac{1}{\sqrt{\Delta y^2 + \Delta \phi^2}} (-\Delta \phi, \Delta y)=(- \sin \beta,  \cos \beta)
, 
\end{align}
where $\Delta y= y_b-y_a$ and $\Delta \phi= \phi_b-\phi_a$, as depicted in Fig.~\ref{rap-azim}. The angle $\beta$ has been introduced for future convenience. 
We now introduce two new observables that are defined as the projections of the pull vector in the two directions identified by the unit vectors above:
\begin{align}\label{pull-proj}
t_{\parallel}&= |\vec{t}\cdot \hat{n}_\parallel | \quad \text{and} \quad
t_{\perp}= |\vec{t}\cdot \hat{n}_\perp |.
\end{align}
We will come back to the role of the absolute value in the expressions above in Section~\ref{sec:asym}.
Furthermore, we note that the magnitude of the pull vector can be expressed as
\begin{equation}
t=|\vec{t}|=  \left | \frac{1}{p_{ta}}\sum_{i\in J}p_{ti} |\vec{r}_i|^2 \hat r_i \right| = \sqrt{t_{\parallel}^2+t_{\perp}^2},
\end{equation}
while the pull angle can be written as
\begin{equation}
\theta_p= \cos^{-1} \frac{\vec{t}\cdot \hat{n}_\parallel}{t}.
\end{equation}
It is easy to check that the pull magnitude $t$ and the two projections $t_{\parallel}$ and $t_{\perp}$ are IRC safe observables. However, this property is lost when considering the pull angle, essentially because $\theta_p$ does not vanish in the presence of a single soft emission because the ratio $t_\parallel / t$ is undetermined. We also note that the projections of pull introduced here share some similarities with jet dipolarity~\cite{Hook:2011cq}.

Our first aim in what follows is to obtain all-order predictions for the above safe observables at next-to-leading logarithmic accuracy. In Ref.~\cite{Larkoski:2019urm} we have already performed a resummed calculation for the pull magnitude $t$, which then played the role of the IRC safe companion observable in the Sudakov safe calculation for $\theta_p$. However, in that calculation we have resorted to the collinear limit. Here, we want to relax this approximation and also consider contributions from soft emissions at wide angle, expressed as a power series in the jet radius $R$.
Crucially, soft radiation at wide angle depends on the number of hard partonic legs present in the processes and on their kinematic configurations. Therefore, in order to perform our calculation we have to choose a process (or a class of processes) and fix the number of coloured legs. 

In this paper we concentrate on measuring pull on one of the two jets originating from the hadronic, i.e.\ $b \bar b$, decay of a Higgs boson, while taking the other jet as reference. More specifically, we focus on the inclusive production of the Higgs together with a $Z$ boson.
We point out that, as suggested in the original publication, pull can provide a valuable handle in distinguishing the above production of a Higgs boson from the dominant QCD background (specifically $g \to b \bar b$). 
Furthermore, this measurement can be also performed in the boosted regime, where the decay products are reconstructed into a single two-pronged jet. In this case, jet pull can be measured on one of the subjets.

We also advocate measuring jet pull in other Standard Model contexts.
Measurements of the pull angle have been carried out by the D$\emptyset$ collaboration at the Tevatron~\cite{Abazov:2011vh} and by the ATLAS collaboration at the LHC~\cite{Aad:2015lxa,Aaboud:2018ibj} (in their most recent analysis the ATLAS collaboration also measured the pull magnitude) in events featuring the production of a top and of an anti-top. 
The rich phenomenology of top decay allows for measuring jet pull in a singlet decay by looking, for instance, at the decay of the $W$ boson but also enables one to study more intricate colour correlations, by measuring the pull between one of the $b$-jets and the incoming beam. 
Another interesting channel to consider is $Z$+jet production. 
This channel offers several interesting possibilities in the context of colour-flow measurements.
For instance, by looking at the substructure of QCD jets, one can explore colour flow in higher-dimensional colour representation, see e.g.~\cite{Bao:2019usu}. 
On the other hand, one can look at the hadronic decay of the $Z$ boson and measure colour flow between two jets (or subjets, if considering the boosted regime) originating from a colour singlet. This situation is very much analogous to what we discuss in this current work, but it features a higher rate at the LHC. 
Studies of colour flow in this context would provide a useful testing ground for an even more interesting Higgs and new physics programme.

\section{Pull distributions at next-to-leading logarithmic accuracy}\label{sec:theory}
In this section we provide all-order calculations that resum large logarithms up to next-to-leading logarithmic accuracy (NLL) for the IRC safe projections of the pull vector considered in this study, namely $t$, $\tpp$ and $\tpa$. These calculations can also be used as input for the Sudakov-safe determination of $\theta_p$.

\subsection{Collinear emissions}

The NLL resummation of the pull vector in the collinear limit, was already performed in Ref.~\cite{Larkoski:2019urm}.
The all-order expression can be easily arrived at by noting that the pull vector is additive and recoil-free at leading power, essentially because of the quadratic dependence  on $|\vec{r}_i|$ of Eq.~(\ref{pull-def})~\footnote{It would be interesting to study observables with a generalised $|\vec{r}_i|^\alpha$ dependence, perhaps employing different recombination schemes in the jet algorithm, such as winner-take-all~\cite{Larkoski:2014uqa}, in order to maintain the recoil-free property. We thank Jesse Thaler for pointing this out.}.
Furthermore, despite the fact that we have in mind to measure jet pull on the hadronic decay products of a Higgs boson, we note that in the collinear limit the resummed cross section is universal and does not depend on the event surrounding the jet we are measuring.
The resummed expression for the pull magnitude can be directly calculated from an infinite sum of emissions of energy fraction $z_i$ and (small) emission angles~$\theta_i\ll R$
\begin{align}\label{eq:t-resum}
\frac{1}{\sigma}\frac{d \sigma}{d t} &= \exp\left[
-\int_0^{R^2}\frac{d\theta^2}{\theta^2}\int_0^1 dz \int_0^{2\pi}\frac{d\phi}{2\pi}\frac{\alpha_s(z \theta p_{ta})}{2\pi}P_{gq}(z)
\right]
\nonumber \\
&
\hspace{1cm}
\times
\left[
\phantom{\delta\left(t- \sqrt{ \left(\sum_{i=1}^n z_i \theta_i^2\cos\phi_i\right)^2+ \left(\sum_{i=1}^n z_i \theta_i^2\sin\phi_i\right)^2}  \right) }
\hspace{-8.4cm}
\sum_{n=0}^\infty\frac{1}{n!} \prod_{i=1}^n \int_0^{R^2}\frac{d\theta_i^2}{\theta_i^2}\int_0^1 dz_i \int_0^{2\pi}\frac{d\phi_i}{2\pi}\frac{\alpha_s (z_i \theta_i p_{ta})}{2\pi}P_{gq}(z_i)\right.\nonumber\\
&
\hspace{2cm}
\left.
\times\, \delta\left(t- \sqrt{ \left(\sum_{i=1}^n z_i \theta_i^2\cos\phi_i\right)^2+ \left(\sum_{i=1}^n z_i \theta_i^2\sin\phi_i\right)^2}  \right) 
\phantom{int_0^{2\pi}}
\hspace{-0.8cm}\right],
\end{align}
where $R$ is the radius of the jet we are measuring pull on. For definiteness, we are going to define jets using the anti-$k_t$ algorithm~\cite{Cacciari:2008gp}.
 The function $P_{gq}=C_F\frac{1+(1-z)^2}{z}$ represents the collinear splitting probability of a quark into a quark and a gluon and appears in the resummation formula because at NLL the parton originating a jet in $H\to b\bar b$ decay is always a quark. A more refined calculation, namely NLL$'$, would also account for the relative $\mathcal{O}(\as)$ probability of measuring pull on a gluon-initiated jets and would therefore would also feature the splitting probabilities $P_{gg}$ and $P_{qg}$. 
Furthermore, note that the argument of the running coupling, which must be evaluated at two-loop accuracy, is the transverse momentum of the emission relative to the hard quark.
As already noticed in Ref.~\cite{Larkoski:2019urm}, the structure of the resummed results is akin to the well-known transverse-momentum resummation, e.g.~\cite{Parisi:1979se,Collins:1984kg}, and consequently the sum over the emissions can be performed explicitly in the conjugate space of Fourier-Hankel moments:
\begin{align}\label{eq:coll-t-res-final}
\frac{1}{\sigma}\frac{d \sigma}{d t}&= \int_0^\infty db\, (bt)J_0(bt)e^{-2C_F \rad_c(b)}, 
\end{align}
where $J_0(x)$ is the Bessel function and $\rad_c(b)$ is the collinear radiator, which, at this accuracy, depends exclusively on the magnitude of the Fourier conjugate vector $b=|\vec b|$:
\begin{align}\label{coll-rad}
\rad_c(b) = \int_0^{R^2}\frac{d\theta^2}{\theta^2}\int_0^1 dz \frac{\alpha_s(z \theta p_{ta})}{2\pi}\frac{P_{gq}(z)}{2C_F}\Theta \left( z \theta^2 -{\bar b}^{-1}\right),
\end{align}
with $\bar b = b \frac{e^{\gamma_E}}{2}$. Explicit expressions for the NLL radiator will be reported in Section~\ref{sec:resummed-formulae}.

The projections of the pull vector we are interested in can be found following the same steps. We have
\begin{align}\label{eq:tpp-resum}
\frac{1}{\sigma}\frac{d \sigma}{d \tpp } &= \exp\left[
-\int_0^{R^2}\frac{d\theta^2}{\theta^2}\int_0^1 dz \int_0^{2\pi}\frac{d\phi}{2\pi}\frac{\alpha_s(z \theta p_{ta})}{2\pi}P_{gq}(z)
\right]
\nonumber \\
&
\hspace{1cm}
\times
\left[
\sum_{n=0}^\infty\frac{1}{n!} \prod_{i=1}^n \int_0^{R^2}\frac{d\theta_i^2}{\theta_i^2}\int_0^1 dz_i \int_0^{2\pi}\frac{d\phi_i}{2\pi}\frac{\alpha_s (z_i \theta_i p_{ta})}{2\pi}P_{gq}(z_i)\right.\nonumber\\
&
\hspace{2cm}
\left.
\times\, 
\delta\left(\tpp- \left | \sum_{i=0}^n\left(-z_i \theta_i^2\cos\phi_i \sin \beta +  z_i \theta_i^2\sin\phi_i\cos \beta \right) \right|\right)\right],
\end{align}
where the $\delta$ function comes from the definition of the observable $\tpp$ in Eq.~(\ref{pull-proj}). Note that in this case such constraint involves a one-dimensional sum, while the analogous term in the pull magnitude distribution, Eq~(\ref{eq:t-resum}), involved a vector sum.
This situation presents strong similarities with the resummation of equivalent variables in the context of transverse-momentum resummation, such as $a_T$ and $\phi^*$~\cite{Banfi:2009dy,Banfi:2011dx}.
Thus, as in that case, the all-order sum can performed in a conjugate Fourier space. We obtain
\begin{align}\label{eq:coll-tpp-res-final} 
\frac{1}{\sigma}\frac{d \sigma}{d \tpp}&= \frac{2}{\pi }\int_0^\infty db\, \cos(b\tpp)e^{- 2C_F \rad_c(b)},
\end{align}
where the radiator in $b$ space is the same as the one obtained for the pull magnitude, Eq.~(\ref{coll-rad})
Finally, we find that, at this accuracy,  the $\tpa$ and $\tpp$ distributions share an identical collinear structure:
\begin{align}\label{eq:coll-res-final2}
\frac{1}{\sigma}\frac{d \sigma}{d \tpa}&= \frac{2}{\pi }\int_0^\infty db\, \cos(b\tpa)e^{- 2C_F\rad_c(b)}.
\end{align}

\subsection{Soft emissions at wide angle}
We now focus our attention on the effect that soft emissions at wide angle have to the pull distributions. These contributions first appear at NLL and from general considerations we expect them to be suppressed in the small jet radius limit. However, unlike collinear radiation discussed above, the explicit form of soft contributions depends on the underlying hard processes we are considering. Physically, this comes about because soft gluons can attach to any hard parton, resulting in a potentially complicated pattern of colour correlations. 
In our current study, the situation is not too complicated because we are focusing on measuring pull on jets originating from a colour-singlet, while the colour structure is much richer when considering jets originating from higher-dimensional colour representations~\cite{Bao:2019usu}. 
In particular, the hard process we are considering at Born level is
\begin{equation}
 q \bar q \to H(\to b \bar b) \; Z (\to l^+ l^-).
\end{equation}

The soft contribution to the NLL radiator can be written as the sum over dipoles that can emit a soft gluon. In our case we only have two dipoles: the one formed by the initial-state partons and the one made up by the two bottom quarks, which we consider massless, therefore we have
\begin{align}\label{soft-radiator}
\mathcal{R}_s= -2{\bf T}_1 \cdot {\bf T}_2 \rad_{12}-2 {\bf T}_a \cdot {\bf T}_b \widetilde{\rad}_{ab},
\end{align}
where $1,2$ refer to the initial state and $a,b$ to the final state. ${\bf T}_i$ are the colour insertion operators and the tilde on the second contribution indicates that we have subtracted the collinear contribution already included in $\rad_c$. Because we are considering final-state jets produced by the decay of a singlet state, the colour algebra is trivial:
\begin{align}\label{colour-alg}
{\bf T}_1 + {\bf T}_2=0 \Rightarrow {\bf T}_1 \cdot {\bf T}_2= -\frac{1}{2} \left({\bf T}_1^2 +{\bf T}_2^2 \right)= - C_F, \nonumber \\
{\bf T}_a + {\bf T}_b=0 \Rightarrow {\bf T}_a \cdot {\bf T}_b= -\frac{1}{2} \left({\bf T}_a^2 +{\bf T}_b^2 \right)= - C_F, \nonumber \\
\end{align}

We start by considering the contribution from the initial-state dipole. Indicating with $p_1$ and $p_2$ the momenta of the incoming quarks and with $k$ the momentum of the soft gluon, we have
\begin{align}\label{1gluon-IRS-begin}
\rad_{12}= \int d k_t k_t d y \frac{d \phi}{2 \pi} \frac{\as(k_t)}{2 \pi} \frac{p_1\cdot p_2}{p_1 \cdot k \; p_2 \cdot k} \Theta_\text{jet} \Theta_\text{pull},
\end{align}
where $\Theta_\text{jet}$ enforces the gluon to be recombined with one of the final-state partons (say parton $a$) to form the jet we are interested in, and $\Theta_\text{pull}$ enforces the gluon contribution to the observable of choice to be above a certain value. 

The above integrals can be easily evaluated by introducing polar coordinates in the rapidity-azimuth plane:
\begin{align}\label{polar-coords}
y-y_a& = r \cos \alpha, \nonumber\\
\phi-\phi_a&= r \sin \alpha.
\end{align}
With this choice of variables, the observables become
\begin{align}\label{1gluon-pull-proj}
t&=|\vec{t}|=z r^2, \nonumber\\
t_{\parallel}&= |\vec{t}\cdot \hat{n}_\parallel |= z r^2 | \cos (\alpha-\beta)|, \nonumber \\
t_{\perp}&= |\vec{t}\cdot \hat{n}_\perp |= z r^2 | \sin (\alpha-\beta)|,
\end{align}
with $z = \frac{k_t}{p_{ta}}$. The angle $\beta$ was introduced in Eq.~(\ref{unit-vectors}).
Note that $\alpha-\beta$ is just the pull angle.

Thus, for the pull magnitude, we obtain
\begin{align}\label{1gluon-IRS-t}
\rad_{12}= \int_0^1  \frac{dz}{z} \frac{\as(z p_{ta})}{\pi}  \int_0^R d r r \int_0^{2 \pi} \frac{d \alpha}{2 \pi} \Theta(z r^2>t)=
R^2 \int_t^1 \frac{dz}{z} \frac{\as(z p_{ta})}{2\pi}+ \dots
\end{align}
where the dots indicate subleading contributions. 
To NLL, the same expression also holds for $t_\parallel$ and $t_\perp$:
\begin{align}\label{1gluon-IRS-begin-final2}
\rad_{12}&= \int_0^1  \frac{dz}{z} \frac{\as(z p_{ta})}{\pi}  \int_0^R d r r \int_0^{2 \pi} \frac{d \alpha}{2 \pi} \Theta(z r^2 |\cos(\alpha-\beta)|>t_\parallel)=
R^2 \int_{t_\parallel}^1 \frac{dz}{z} \frac{\as(z p_{ta})}{2\pi}+ \dots\\
\rad_{12}&= \int_0^1  \frac{dz}{z} \frac{\as(z p_{ta})}{\pi}  \int_0^R d r r \int_0^{2 \pi} \frac{d \alpha}{2 \pi} \Theta(z r^2 |\sin(\alpha-\beta)|>t_\perp)=
R^2 \int_{t_\perp}^1 \frac{dz}{z} \frac{\as(z p_{ta})}{2\pi}+ \dots
\end{align}
where again the dots indicate subleading contributions. 

Thus far we have calculated the soft wide-angle contribution directly in momentum space. This is in principle sufficient at NLL accuracy we are working at. Nevertheless, in order to smoothly combine the soft contribution to the collinear one previously computed, we find convenient to perform the whole resummation in moment ($b$) space. Therefore to NLL we can write the soft contribution from the initial-state dipole as
\begin{align}\label{1gluon-IRS-begin-final}
\rad_{12}&=R^2 \int_{1/\bar{b}}^1 \frac{dz}{z} \frac{\as(z p_{ta})}{2\pi}.
\end{align}

Next we consider soft-wide angle emissions off the final-state $ab$ dipole. 
As in the previous case, we find convenient to express the phase-space integrals in polar coordinates. 
We have
\begin{align}\label{1gluon-FRS-begin}
\rad_{ab}&= \int d k_t k_t d y \frac{d \phi}{2 \pi} \frac{\as(\kappa_{ab})}{2 \pi} \frac{p_a\cdot p_b}{p_a \cdot k \; p_b \cdot k} \Theta_\text{jet} \Theta_\text{pull} \nonumber\\
&= \int_0^1  \frac{dz}{z}  \int_0^R d r  \int_0^{2 \pi} \frac{d \alpha}{2 \pi} \frac{\as(\kappa_{ab})}{2 \pi} 
\left[\frac{2}{r}+ \mathcal{A}(\alpha,\beta)+ \mathcal{B}(\alpha,\beta) r+ \dots \right]
\Theta_\text{pull}
\end{align}
where the argument of the running coupling $\kappa_{ab}^2=\frac{2\; p_a \cdot k \; p_b \cdot k}{p_a\cdot p_b}$ is the transverse momentum of the gluon with respect to the dipole, in the dipole rest frame. 
We calculate this contribution as a power expansion in the jet radius $R$, which corresponds to expanding the integrand in powers of $r$. 
The first contribution within the square brackets is the soft and collinear piece, which we have already accounted for in $\rad_c$. Therefore, we consider 
\begin{align}\label{dipole-expanded}
\widetilde{\rad}_{ab}&=
 \int_0^1  \frac{dz}{z}  \int_0^R d r  \int_0^{2 \pi} \frac{d \alpha}{2 \pi} \frac{\as(\kappa_{ab})}{2 \pi} 
\left[ \mathcal{A}(\alpha,\beta)+ \mathcal{B}(\alpha,\beta) r+ \dots \right]
\Theta_\text{pull}
\end{align}
The first term above, namely $\mathcal{A}$ gives no contribution when we integrate over all possible angles. It would give rise to an $\mathcal{O}(R)$ correction if we impose further angular restrictions. We will come back to this observation in Section~\ref{sec:asym}.  The $\mathcal{B}$ term gives rise to a contribution which is identical in all cases. Therefore, at NLL we have
\begin{align}\label{1gluon-FRS-la}
\widetilde{\rad}_{ab}&= \frac{R^2}{4} \frac{\cosh \Delta y+ \cos \Delta \phi}{\cosh \Delta y- \cos \Delta \phi}\int_{1/\bar{b}}^1 \frac{dz}{z} \frac{\as(z p_{ta})}{2\pi} + \mathcal{O}(R^4).
\end{align}
We remind the reader that explicit expressions for the NLL radiator will be reported in Section~\ref{sec:resummed-formulae}.

\subsection{Non-global logarithms}\label{sec:ngls}

Jet pull is measured on an isolated jet and it is therefore a text-book example of a non-global observable~\cite{Dasgupta:2001sh}.
In this section we investigate the structure of non-global logarithms (NGLs) that affect the different projections of the pull vector. 

We focus on the final-state dipole $ab$ and we consider the double differential distribution in the pull magnitude and pull angle at $\mathcal{O}(\as^2)$.
To calculate the leading non-global logarithmic contribution to the pull vector, it suffices to consider correlated soft gluon emission from the dipole in which the two soft gluons have parametrically separated energies $k_h\gg k_s$, in the phase-space region where the harder gluon lies outside the measured jet, while the second one is inside. The matrix element for this non-global contribution can then be expressed as
\begin{align}\label{ngls-start}
\frac{d^2\sigma^\text{NG}}{dt\, d\theta_p} &= \frac{\alpha_s^2C_F C_A}{16\pi^4}   \int_0^1 \frac{dk_{\perp h}}{k_{\perp h}}\int_{-\infty}^\infty dy_h \int_{-\pi}^\pi d\phi_h  \int_0^1 \frac{dk_{\perp s}}{k_{\perp s}}\int_{-\infty}^\infty dy_s\int_{-\pi}^\pi d\phi_s  \, \frac{2 p_a\cdot p_b}{(p_a\cdot k_h) (p_b\cdot k_h)}\nonumber\\
&
\hspace{3cm}
\times\frac{(p_a\cdot k_h)(p_b\cdot k_s)+(p_a\cdot k_s)(p_b\cdot k_h)-(p_a\cdot p_b)(k_h\cdot k_s)}{(p_a\cdot k_s)(p_b\cdot k_s)(k_h\cdot k_s)}\\
&
\hspace{3cm}
\times \Theta\left(R^2-(y_s-y_a)^2-(\phi_s-\phi_a)^2\right)\Theta\left((y_h-y_a)^2+(\phi_h-\phi_a)^2-R^2\right)\nonumber\\
&
\hspace{3cm}
\times\, \Theta(k_{\perp h}\cosh y_h - k_{\perp s}\cosh y_s)\,\delta\left(t-k_{\perp s}\left((y_s-y_a)^2+(\phi_s-\phi_a)^2\right)\right)\nonumber\\
&
\hspace{3cm}
\times\,\delta\left(\theta_p-\cos^{-1}\frac{(y_s-y_a)\cos\beta +(\phi_s-\phi_a)\sin \beta }{\sqrt{(y_s-y_a)^2+(\phi_s-\phi_a)^2}}\right)
\,.\nonumber
\end{align}
Note that in the expression, the dependence on the perp magnitudes has been pulled out of all of the matrix elements and made explicit.
The integral over $k_{\perp s}$ and  $k_{\perp h}$ can easily performed. Furthermore, for compactness, we can shift the $y$ and $\phi$ coordinates to be measured with respect to the location of jet $a$, i.e.\ without loss of generality we can set $y_a=\phi_a=0$ in Eq.~(\ref{ngls-start}).

From this point, we will start approximating the integrals that remain.  First, we only work to find the leading NGLs for $t\ll 1$.  
Then, we consider the phase-space constraints that remain and we notice that, in the small jet radius limit, we have the following scaling $ y_h\sim  y_s \sim R \ll1$.
Therefore, in the explicit logarithm in the integrals we can simply remove the hyperbolic cosine factors, as their contribution will be purely beyond leading NGL.  Correspondingly, because $R\ll 1$, we can push the bounds of integration on $ \phi_s, \phi_h$ safely to infinity.  The integrals then become
\begin{align}
\frac{d^2\sigma^\text{NG}}{dt\, d\theta_p} &= \frac{\alpha_s^2C_F C_A}{16\pi^4}  \, \frac{1}{t} \int_{-\infty}^\infty d  y_h \int_{-\infty}^\infty d \phi_h  \int_{-\infty}^\infty d y_s\int_{-\infty}^\infty d \phi_s  \, \frac{2 p_a\cdot p_b}{(p_a\cdot k_h) (p_b\cdot k_h)}\\
&
\hspace{2cm}
\times\frac{(p_a\cdot k_h)(p_b\cdot k_s)+(p_a\cdot k_s)(p_b\cdot k_h)-(p_a\cdot p_b)(k_h\cdot k_s)}{(p_a\cdot k_s)(p_b\cdot k_s)(k_h\cdot k_s)}\,\log\frac{ y_s^2+\phi_s^2}{t}\nonumber\\
&
\hspace{2cm}
\times
\, \Theta\left(y_s^2+\phi_s^2-t\right)\, \Theta\left(R^2-y_s^2-\phi_s^2\right)\,\Theta\left(y_h^2+\phi_h^2-R^2\right)\nonumber\\
&
\hspace{2cm}
\times\,\delta\left(\theta_p-\cos^{-1}\frac{y_s \cos\beta +\phi_s \sin \beta }{\sqrt{y_s^2+\phi_s^2}}\right)
\,.\nonumber
\end{align}
Similarly to the one-gluon dipoles previously discussed, the integrals are more easily performed in polar coordinates, see Eq.~(\ref{polar-coords}):
\begin{align}
 y_i& =r_i \cos \gamma_i\,,\nonumber\\
\phi_i&=r_i \sin\gamma_i\,.
\end{align}
Then, the integrals become
\begin{align}
\frac{d^2\sigma^\text{NG}}{dt\, d\theta_p} &= \frac{\alpha_s^2C_F C_A}{16\pi^4}  \, \frac{1}{t} \int_0^\infty dr_h\, r_h \int_0^{2\pi} d\gamma_h  \int_0^\infty dr_s\, r_s\int_0^{2\pi} d\gamma_s  \, \frac{2 p_a\cdot p_b}{(p_a\cdot k_h) (p_b\cdot k_h)}\\
&
\hspace{2cm}
\times\frac{(p_a\cdot k_h)(p_b\cdot k_s)+(p_a\cdot k_s)(p_b\cdot k_h)-(p_a\cdot p_b)(k_h\cdot k_s)}{(p_a\cdot k_s)(p_b\cdot k_s)(k_h\cdot k_s)}\nonumber\\
&
\hspace{2cm}
\times
\,\log\frac{r_s^2}{t}\, \Theta\left(r_s^2-t\right)\, \Theta\left(R-r_s\right)\,\Theta\left(r_h-R\right)\,\delta\left(\theta_p-\gamma_s +\beta \right)
\,.\nonumber
\end{align}
Now, we need to express the soft matrix element in these coordinates.  Additionally, we work in the small jet radius limit, $R\ll1$, and note that the dominant contribution to the NGLs comes from the region of phase space in which $r_s\lesssim r_h \sim R$.  We will thus expand the matrix element to first order in the $R \ll 1$ limit with this identified scaling.  We find
\begin{align}
\frac{d^2\sigma^\text{NG}}{dt\, d\theta_p} &=  \left(\frac{\alpha_s}{2\pi}\right)^2 C_F C_A\,\frac{\pi}{3} \, \frac{\log\frac{R^2}{t}}{t}\biggl(1+\frac{24 (1-\log\, 2)}{\pi^2}R\\
&
\hspace{0.5cm}
\left.\, \times \frac{\sin  \Delta \phi \; \sin(\theta_p+\beta)+\sinh \Delta y \;\cos(\theta_p+\beta)}{\cosh  \Delta y-\cos \Delta \phi}+{\cal O}(R^2)\right)\,.\nonumber
\end{align}
The first term in this expansion is the familiar expression for the narrow jet mass NGL matrix element.
Note that this differs by a factor of $2\pi$ from the familiar expression for the jet mass NGLs; this factor is recovered when $\theta_p$ is integrated over.
Furthermore, if we integrate over the full range for $\theta_p$, then the contribution which is linear in $R$ vanishes, leading to
\begin{equation}\label{ngls-final-as2}
\frac{d \sigma^\text{NG}}{dt} = \left(\frac{\alpha_s}{2\pi}\right)^2 C_F C_A\frac{2\pi^2}{3}  \, \frac{\log\frac{R^2}{t}}{t}+{\cal O}(R^2)\,.
\end{equation}
It is easy to verify that at NLL accuracy the same expression as Eq.~(\ref{ngls-final-as2}) holds for the projections $\tpa$ and $\tpp$.

If we only to retain the leading $R$ term, then resummation of NGLs is analogous as the hemisphere mass originally studied in~\cite{Dasgupta:2001sh}. 
We could, in principle, also include the $\mathcal{O}(R^2)$ corrections, as done in the global part. This would require evaluating the subsequent term in the small-$R$ expansion of Eq.~(\ref{ngls-final-as2}). Furthermore, we would also have to include the NGL contribution from initial-state radiation, as discussed, for instance in Ref.~\cite{Dasgupta:2012hg}, in the context of jet mass distributions. 
We leave this study for future work and, in this current study, we limit ourselves to a numerical estimate of this effect, as detailed in Section~\ref{sec:numerics}.

\subsection{Resummed results}\label{sec:resummed-formulae}

We are now in a position to collect all the results derived so far and obtain a NLL resummed prediction for the safe projections of the pull vector we are considering. 
The all-order differential distribution can be written as:
\begin{align}\label{final-res-expr}
\frac{1}{\sigma}\frac{d \sigma}{d v} &=\int_0^\infty d b\,  \mathcal{F}_v(b v) e^{-C_F \mathcal{R}(b) }\mathcal{S}^\text{NG}(b),
\end{align}
with
\begin{align}
\mathcal{F}_v(x)=
\begin{cases}
x J_0(x), & \text{if} \quad v=t,\\
\frac{2}{\pi}\cos (x), &\text{if} \quad v=\tpa, \, \tpp.
\end{cases}
\end{align}
The resummed exponent $\mathcal{R}$ can be written in terms of leading (second line) and next-to-leading (third to fifth lines) contributions: 
\begin{align}\label{radiator}
\mathcal{R}&=2\mathcal{R}_c+2\widetilde{\mathcal{R}}_{ab}+2\mathcal{R}_{12}  \nonumber \\ 
&=
 \frac{\left(1-2 \lambda \right ) 
\log \left(1-2\lambda \right)-2 \left ( 1-\lambda \right ) \log \left
  (1-\lambda \right )}{2 \pi \as \beta_0^2}
 \nonumber \\
&+
 \frac{ B_q }{ \pi \beta_0} \log \left ( 1-\lambda \right ) 
 + \frac{ K}{4 \pi^2 \beta_0^2} \left [2 \log \left 
(1-\lambda \right ) - \log \left (1-2 \lambda \right )\right ]  \nonumber\\ &
  +\frac{ \beta_1}{2 \pi \beta_0^3} \left [ \log \left (1-2\lambda \right )-2 \log 
\left (1-\lambda \right ) + \frac{1}{2} \log^2 \left (1- 2 \lambda \right ) 
- \log^2 \left (1-\lambda \right ) \right ]
\nonumber\\ 
&+\frac{1}{\pi \beta_0}\log \frac{p_{ta} R}{\mu_R}\left[\log(1-2 \lambda) -2 \log(1-\lambda)\right]
- \frac{R^2}{8 \pi \beta_0} \left[ 4+ \frac{\cosh \Delta y +\cos \Delta \phi}{\cosh \Delta y -\cos \Delta \phi} \right]\log (1-2 \lambda) \nonumber \\&+\mathcal{O}(R^4),
\end{align}
with $\lambda= \as \beta_0 \log (\bar{b} R^2)$\footnote{Strictly speaking the jet radius dependence in argument of the logarithms only appears at this order in the soft-collinear contributions. However, we find that including it in the whole radiator, leads to better numerical stability. The difference between these choices is beyond NLL accuracy.} and $\as=\as(\mu_R)$, where $\mu_R$ is the renormalisation scale, which we can vary around the hard scale $p_{ta}$ in order to assess missing higher-order corrections.
In the above results the $\beta$ function coefficients 
$\beta_0$ and $\beta_1$ are defined as
\begin{equation}
\beta_0 = \frac{11 C_A - 2 n_f }{12 \pi}, \qquad \beta_1 = \frac{17 C_A^2 - 5 C_A n_f -3 C_F n_f}{24 \pi^2}\,,
\end{equation}
and
\begin{equation}
B_q=\frac{3}{4}, \qquad K = C_A \left (\frac{67}{18}- \frac{\pi^2}{6} \right ) - \frac{5}{9} n_f\,.
\end{equation}

Finally, as already mentioned, in the small-$R$ limit, the non-global contribution can be taken equal to the hemisphere case. The resummation of NGLs can be performed in the large-$N_c$ limit exploiting a dipole cascade picture. We make use of the following parametrisation~\cite{Dasgupta:2001sh}: 
\begin{equation}\label{ngl-res}
\mathcal{S}^\text{NG}=\exp \left[-C_F C_A\frac{\pi^2}{3}\frac{1+(a \tau)^2}{1+(b \tau)^c} \tau^2 \right],
\end{equation}
with $\tau=-\frac{1}{4 \pi \beta_0} \log(1- 2 \lambda)$, with $a=0.85 C_A, b=0.86 C_A$, and $c=1.33$.

Finally, we note that the above results are valid for jets defined with the anti-$k_t$ algorithm, which acts as a perfect cone in the soft limit~\cite{Cacciari:2008gp}. Had we use a different clustering measure, such as Cambridge/Aachen~\cite{Dokshitzer:1997in,Wobisch:1998wt} or the $k_t$-algorithm~\cite{Catani:1993hr,Ellis:1993tq}, nontrivial clustering logarithms would have modified both the global and non-global contributions to the resummed exponent~\cite{Banfi:2005gj,Delenda:2006nf,Banfi:2010pa}.

\section{Towards phenomenology}\label{sec:pheno}
In the previous section, we have discussed all the theoretical ingredients that go into a NLL calculation for the jet pull projections considered in this paper. We now turn our attention towards some preliminary phenomenological studies. After discussing a simple model of non-perturbative corrections due to the hadronisation process, we move to compare our resummed results to the one obtained by a general purpose Monte Carlo event generator. While doing so, we also discuss the numerical impact of the various contributions that we have computed thus far. 
We postpone a more detailed phenomenological study, which would also include matching to fixed-order calculations, to future work and we look forward to comparison of our predictions to future experimental measurements. 

\subsection{Non-perturbative corrections}\label{sec:non-pert}

Because the pull vector is both an additive observable and recoil-free, corrections due to non-perturbative physics and hadronisation can be modelled by a shape function \cite{Korchemsky:1999kt,Korchemsky:2000kp,Bosch:2004th,Hoang:2007vb,Ligeti:2008ac}.  This shape function is then convolved with the perturbative distribution to produce a non-perturbative distribution. The shape function depends on a dimensionful relative transverse-momentum scale $\epsilon$, and it has most of its support around $\epsilon = \Lambda_\text{QCD}$, the QCD scale.  The shape function for the pull vector also has non-trivial azimuthal angle dependence, because non-perturbative emissions will be emitted in a preferential direction according to the dipole configuration. 

In this section, we will construct a shape function for the pull vector, assuming that it exclusively has support at $\epsilon = \Lambda_\text{QCD}$.  Further, we will assume that the dominant non-perturbative emission lies exactly at the boundary of the jet on which we measure the pull vector, and its azimuthal distribution about the jet axis is uniform.  We will see that a non-uniform distribution of the pull vector is generated by a preferential emission of higher-energy non-perturbative emissions at small values of the pull angle.

To construct the shape function with these restrictions, we first note that the scale $\epsilon$ for an emission from a dipole with ends defined by the light-like directions $p_a$ and $p_b$ is
\begin{align}
\epsilon = \Lambda_\text{QCD}= \sqrt{(k\cdot p_a)(k\cdot p_b)}\,,
\end{align}
where $k$ is the four-momentum of the non-perturbative emission.  The pull vector depends on the momentum transverse to the beam axis, $k_t$, and its value is constrained by the non-perturbative scale.  Expressing the momentum $k$ as
\begin{align}
k &= k_t(\cosh y,\cos\phi,\sin\phi,\sinh y)\,,
\end{align}
we can express $k_t$ as
\begin{align}
k_t = \frac{\Lambda_\text{QCD}}{\left(\cosh(y-y_a)-\cos(\phi-\phi_a)\right)^{1/2}\left(\cosh(y-y_b)-\cos(\phi-\phi_b)\right)^{1/2}}\,.
\end{align}
Now, we expand this expression to second order in the jet radius $R$, fixing the angle between the non-perturbative emission and the jet axis $n_a$ to be $R$:
\begin{align}
R^2 = (y - y_a)^2+(\phi - \phi_a)^2\,.
\end{align}
We find
\begin{align}
k_t = \frac{2\Lambda_\text{QCD}}{R}\frac{\sqrt{p_{ta}p_{tb}}}{m_H}+2\Lambda_\text{QCD}\frac{(p_{ta}p_{tb})^{3/2}}{m_H^3}\left[
\cos(\varphi+\beta)\sinh \Delta y+\sin(\varphi+\beta)\sin\Delta\phi
\right] + {\cal O}(R)\,.
\end{align}
The relative rapidity $\Delta y$, azimuth $\Delta \phi$, and angle $\beta$ were defined in \Sec{sec:obs}.  The azimuthal angle $\varphi$ defines the angle about the jet axis $p_a$ with respect to $p_b$.  Finally, we have introduced the transverse momentum of the ends of the dipole $p_{ta}$ and $p_{tb}$ and note that they are constrained by the Higgs mass:
\begin{align}
m_H^2 = 2p_{ta}p_{tb}(\cosh\Delta y - \cos\Delta\phi)\,.
\end{align}

With this construction, the shape function for the non-perturbative $k_t$ and azimuthal angle $\varphi$ is
\begin{align}
&F(k_t,\varphi) \\
&= \frac{1}{2\pi}\delta\left(
k_t-\frac{2\Lambda_\text{QCD}}{R}\frac{\sqrt{p_{ta}p_{tb}}}{m_H}-2\Lambda_\text{QCD}\frac{(p_{ta}p_{tb})^{3/2}}{m_H^3}\left[
\cos(\varphi+\beta)\sinh \Delta y+\sin(\varphi+\beta)\sin\Delta\phi
\right]
\right)\,.\nonumber
\end{align}
Given the perturbative pull vector distribution $\frac{1}{\sigma} \frac{d^2 \sigma^\text{pert}}{d \vec{t}^2}$, we now want to find the non-perturbative pull vector distribution $\frac{1}{\sigma} \frac{d^2 \sigma^\text{np}}{d \vec{t}^2}$ through convolution with the shape function. The contribution to pull from the non-perturbative emission that we identified in the rest frame of the Higgs boson will be
\begin{equation}
\vec t_\text{np}(k_t, \varphi) = \frac{k_t R^2}{p_{ta}}\,(\cos\varphi,\sin\varphi)\,.
\end{equation}
It then follows that the non-perturbative distribution of the pull vector is
\begin{align}\label{eq:NP-2D}
\frac{d^2 \sigma^\text{np}}{d \vec{t}^{\;2}}&= \int_0^\infty dk_t \int_0^{2\pi}d\varphi\, F(k_t,\varphi)\, \frac{d^2 \sigma^\text{pert}}{d \vec{t}^{\;2}}\left(
\vec t - \vec t_\text{np}(k_t, \varphi)
\right)\nonumber\\&
=\int_0^{2\pi}\frac{d\varphi}{2\pi}  \,\frac{d^2 \sigma^\text{pert}}{d \vec{t}^{\;2}}\left(
\vec t -  \vec t_\text{np}(k_t, \varphi)
\right)\,,
\end{align}
where we leave the dependence on the non-perturbative transverse momentum $k_t$ implicit.

In order to understand the behaviour of the leading non-perturbative corrections, we expand the above expression in powers of $\Lambda_\text{QCD}$.
Furthermore, we note that because of the particular choice of the reference frame we have used in this section, $\varphi=0$ corresponds to the line joining the two jet centres.
Thus, we obtain
\begin{align}\label{eq:NP-2D-exp}
&\frac{d^2 \sigma^\text{np}}{d \tpa d \tpp}=
\frac{d^2 \sigma^\text{pert}}{d \tpa d \tpp}-\int_0^{2\pi}\frac{d\varphi}{2\pi} \,
 \vec t_\text{np}(k_t, \varphi) \cdot {\bf} \nabla \left(\frac{d^2 \sigma^\text{pert}}{d \tpa d \tpp}\right)+ \mathcal{O}\left(\frac{\Lambda_\text{QCD}^2}{m_H^2} \right)\\&
 =\Bigg[ 1-\frac{\Lambda_\text{QCD}R^2\sqrt{p_{ta}p_{tb}^3}}{m_H^3\sqrt{\Delta y^2+\Delta\phi^2}}\Big(
\left(\Delta y\sinh \Delta y+\Delta\phi\sin\Delta\phi\right)\frac{\partial}{\partial \tpa }\nonumber \\ &\quad +\left(\Delta y\sin \Delta\phi-\Delta\phi\sinh\Delta y\right)\frac{\partial}{\partial \tpp }
\Big)  \Bigg]\frac{d^2 \sigma^\text{pert}}{d \tpa d \tpp }.\nonumber
\end{align}
Because of the derivative dependence in this non-perturbative correction, its effect can be included to lowest order in both $\Lambda_\text{QCD}$ and $\alpha_s$ with a shift of the appropriate argument of the perturbative cross section.  For the cross sections of $\tpa$ and $\tpp$ individually, we have
\begin{align}\label{np-shift-tpa-tpp}
\frac{d \sigma^\text{np}}{d \tpa} &=\frac{d \sigma^\text{pert}}{d \tpa }\left(
\tpa  -\frac{\Lambda_\text{QCD}R^2\sqrt{p_{ta}p_{tb}^3}}{m_H^3\sqrt{\Delta y^2+\Delta\phi^2}}\left(\Delta y\sinh \Delta y+\Delta\phi\sin\Delta\phi\right)\right)+{\cal O}(\Lambda_\text{QCD}^2,\alpha_s)\,,\\
\frac{d \sigma^\text{np}}{d \tpp} &=\frac{d \sigma^\text{pert}}{d \tpp }\left(
\tpp  -\frac{\Lambda_\text{QCD}R^2\sqrt{p_{ta}p_{tb}^3}}{m_H^3\sqrt{\Delta y^2+\Delta\phi^2}}\left(\Delta y\sin \Delta\phi-\Delta\phi\sinh\Delta y\right)\right)+{\cal O}(\Lambda_\text{QCD}^2,\alpha_s)\,.
\end{align}
The leading non-perturbative correction to the magnitude of the pull vector $t$ can be found by exploiting its relationship to $\tpa$ and $\tpp$:
\begin{align}
t = \sqrt{\tpa^2+\tpp^2}\,.
\end{align} 
Then, we have that the pull magnitude distribution becomes
\begin{align}\label{np-shift-t}
\frac{d \sigma^\text{np}}{d t} = \frac{d \sigma^\text{pert}}{d t} \left(t-\frac{\Lambda_\text{QCD}R^2\sqrt{p_{ta}p_{tb}^3}}{m_H^3}\sqrt{\sinh^2\Delta y + \sin^2\Delta\phi}\right)+{\cal O}(\Lambda_\text{QCD}^2,\alpha_s)\,.
\end{align}

\subsection{Numerical studies}\label{sec:numerics}
\begin{figure}[t!]
\centering
\includegraphics[width=0.49 \textwidth]{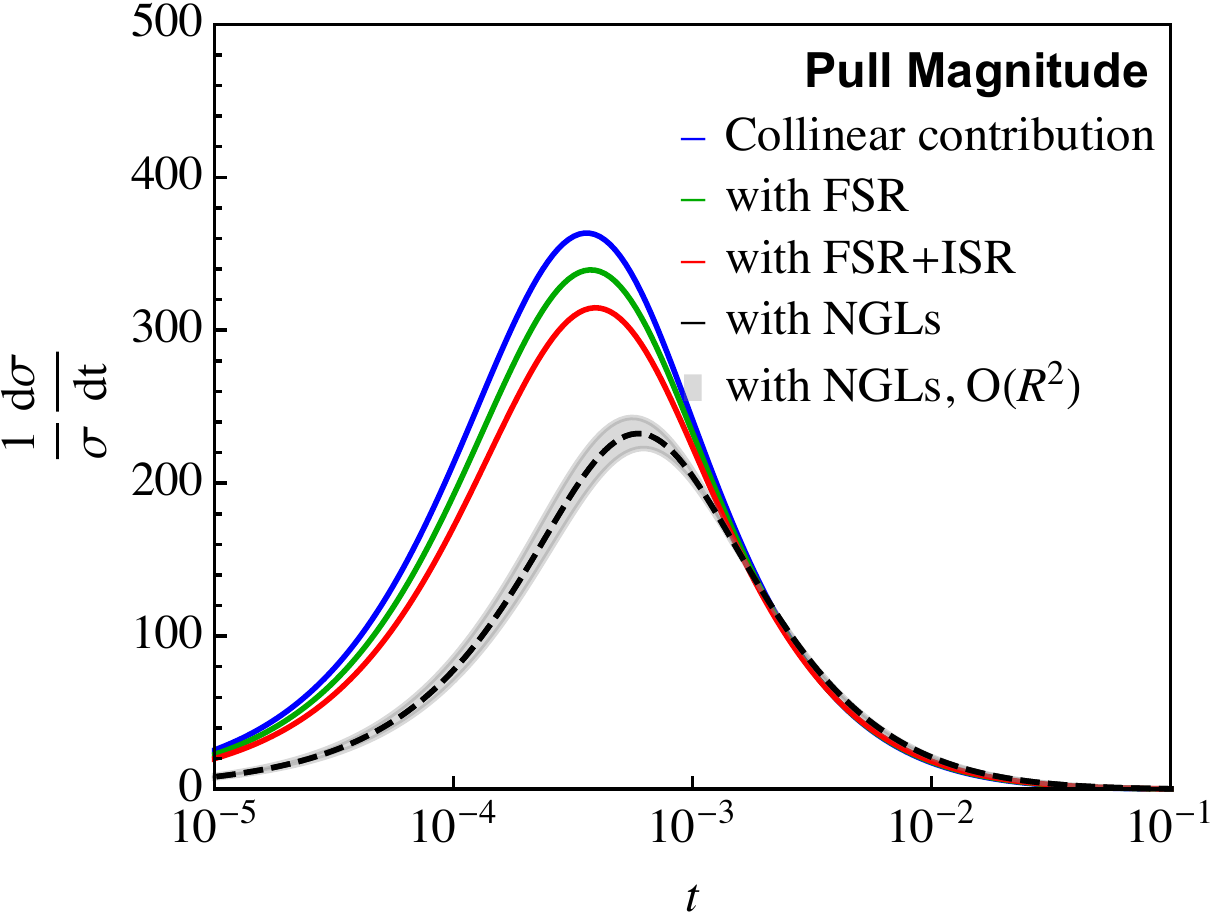}
\includegraphics[width=0.49 \textwidth]{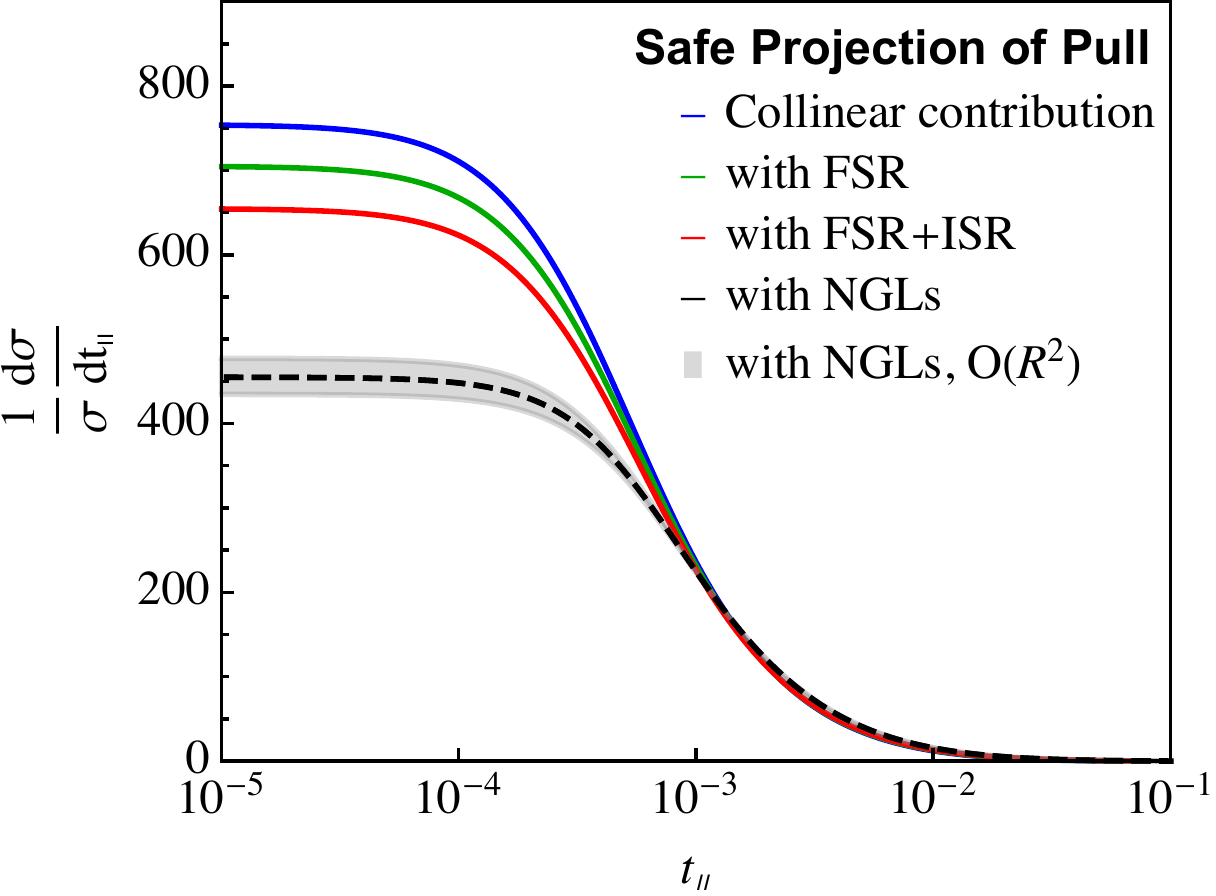}
\caption{Impact of the different contributions to all-order next-to-leading logarithmic resummation of the pull magnitude (left) and the safe projection $\tpa$ (right). Soft gluon contributions at wide angle are included as an expansion in the jet radius $R$ through $\mathcal{O}(R^2)$, while the non-global logarithmic contribution is accounted for at $\mathcal{O}(R^0)$. At this accuracy the distribution of the orthogonal projection $\tpp$ is identical to $\tpa$.}
\label{theory-contributions} 
\end{figure}

We are now ready to perform some phenomenological studies of our results. 
From a technical point of view, we note that the integral over the Fourier variable $b$ which appears in the resummation formula, e.g.\ Eq.~(\ref{final-res-expr}), is ill-defined both at small and large $b$. The bad behaviour at small $b$, which corresponds to large values of the observables, is beyond the jurisdiction of the all-order calculations and it contributes to a region that would be dominated by fixed-order matrix elements. In order to address this issue, we adopt the standard procedure of $Q_T$ resummation~\cite{Bozzi:2005wk} and we shift the argument of the logarithm in $b$-space by unity, i.e. $\log \bar b R^2 \to \log (1+ \bar b R^2)$. 
The resummed exponent is also ill-defined at large $b$ because of the presence of the QCD Landau pole which appears at $\lambda =\frac{1}{2}$. 
We circumvent this issue by further substituting the dependence on the variable $b$ in the resummed exponent with the so-called $b^*$ variable~\cite{Collins:1984kg}
\begin{equation}
b^{*}=\frac{b}{\sqrt{1+\frac{b^{2}}{b_\text{max}^{2}}}},
\end{equation}
where $b_\text{max}$ is chosen in the vicinity of the Landau pole. 
Because $b^*\simeq b$ when $b\ll b_\text{max}$, the perturbative behaviour is unchanged, while the $b$ dependence of the resummed exponent is frozen as $b$ approaches the non-perturbative region, providing us with a prescription to deal with the Landau singularity. 

We start by assessing the numerical impact of the different contributions that are included in our resummed results, namely collinear emissions, final-stare radiation (FSR), i.e.\ the $\mathcal{O}(R^2)$ contribution arising from the final-state dipole, initial-state radiation (ISR), and non-global logarithms. The results are show in Fig.~\ref{theory-contributions}, on the left for the pull magnitude distribution and on the left for the $\tpa$ distribution (at NLL this is the same as $\tpp$). 
The plots are for a representative phase-space point: $\Delta y=1$, $\Delta \phi=\frac{\pi}{6}$ and $p_{ta}=p_{tb}=\frac{m_H}{\sqrt{2(\cosh \Delta y-\cos \Delta \phi})}\simeq 110$~GeV, which corresponds to a symmetric decay of the Higgs boson. 
We note that the collinear approximation describes the two distributions well, down to values of the observables $\sim 10^{-3}$. Below that, in the Sudakov region, the impact of soft-emissions at large angle becomes sizeable. However, we note that finite $R$ corrections, which characterise FSR and ISR are not very large, due to the smallness of the jet radius parameter $R=0.4$, employed in this study. Perhaps surprising is the relatively large contribution due to non-global logarithms. 
This last contribution is shown with an uncertainty band which aims to probe the impact of $\mathcal{O}(R^2)$ corrections to the non-global contribution, which is not included here. The band is constructed by rescaling the  $\mathcal{O}(R^0)$ coefficient by the factor $(1+a R^2)$ and by varying $-1\le a \le 1$. We note that this uncertainty is not large, due to the relatively small value of the jet radius employed here.

By comparing the two distributions, $t$ and $\tpa$, we note that the former exhibits a Sudakov peak, while the latter appears to develop a plateau for $\tpa< 10^{-4}$. This behaviour is completely analogous to what is found when looking at $Q_T$ and $a_T$/$\phi^*$ distributions~\cite{Banfi:2011dx}.
Small values of $t$ or $\tpa$ can be obtained by soft/collinear emissions or by kinematical cancellations and the behaviour of $\tpa$ signals the fact that kinematical cancellation is the dominant mechanism and prevents the formation of the Sudakov peak, as opposed to what happens with $t$.

\begin{figure}[t!]
\centering
\includegraphics[width=0.49 \textwidth]{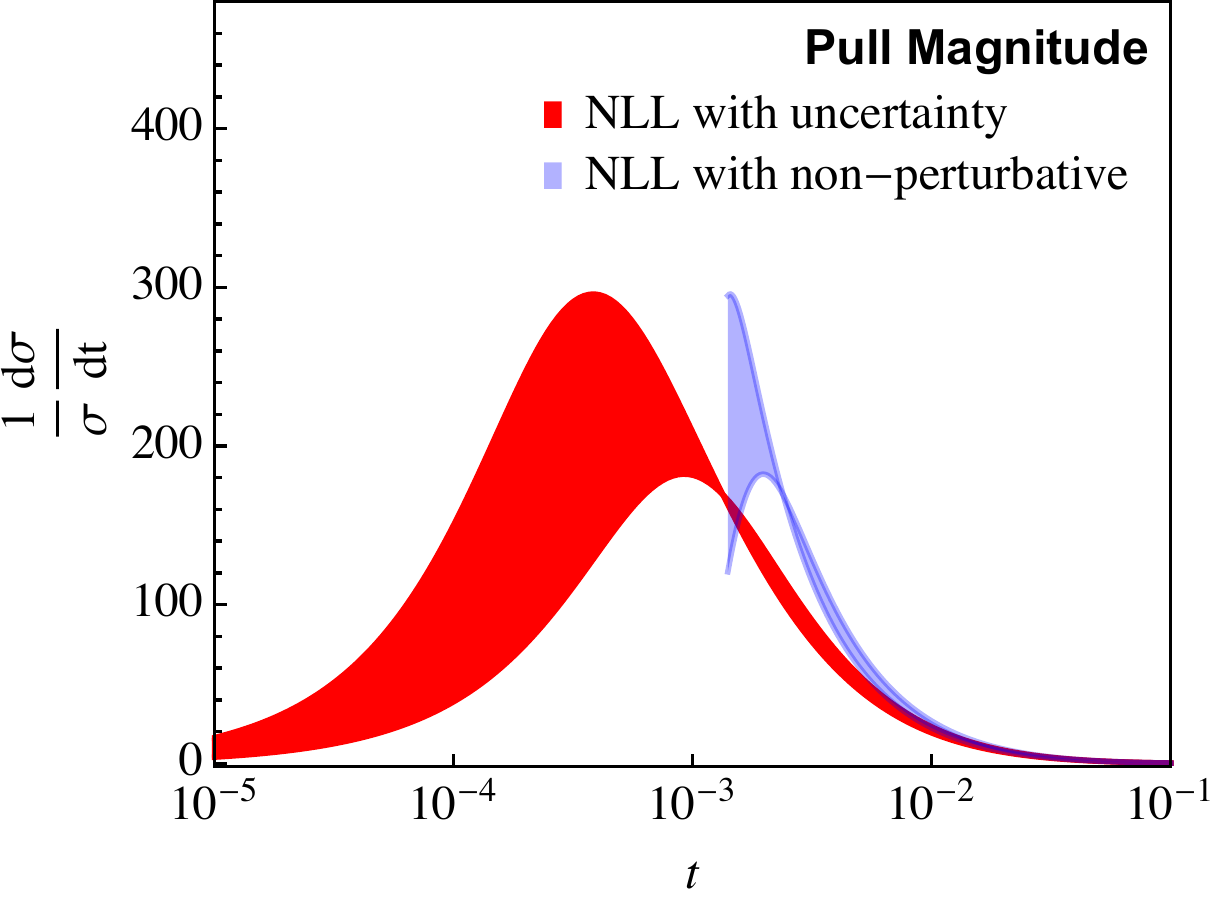}
\includegraphics[width=0.49 \textwidth]{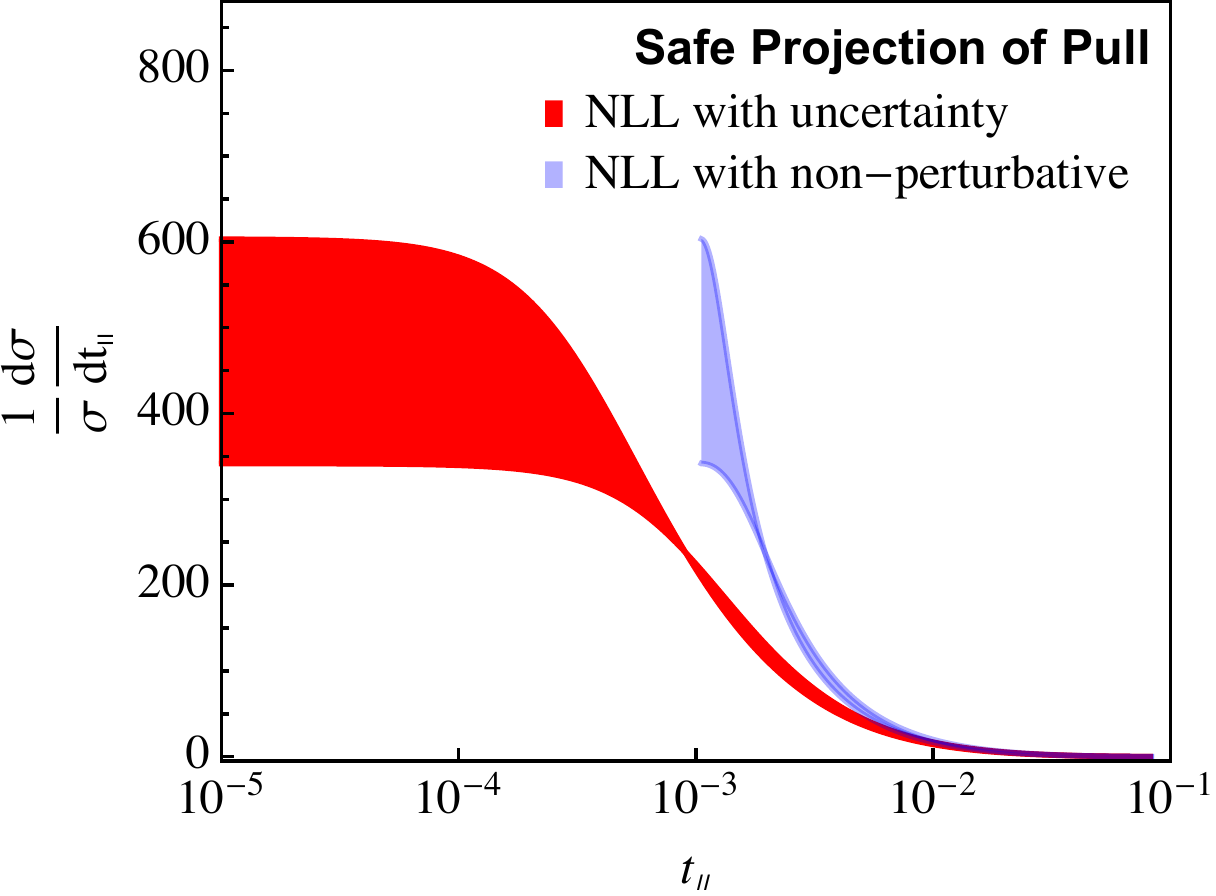}
\caption{Plots of the NLL predictions for $t$ (left) and $\tpa$ (right), together with an estimate of the theoretical uncertainties, obtained by varying the renormalisation scale. The plots also show how the curve is modify once the shift due to non-perturbative corrections is applied.}
\label{theory-uncertainties} 
\end{figure}

Next, in Fig.~\ref{theory-uncertainties} we show our final NLL predictions for $t$ (left) and $\tpa$ (right), with an estimate of the perturbative uncertainty, which we obtain by varying the renormalisation scale in the range $\frac{p_t}{2}\le \mu_R\le 2 p_t$. Furthermore, we also show the NLL calculation supplemented by our estimate of non-perturbative contributions due to the hadronisation process, i.e.\ Eqs.~(\ref{np-shift-tpa-tpp}) and~(\ref{np-shift-t}), using $\Lambda_\text{QCD}=1$~GeV.
We note that because of the $R^2$ coefficient, the size of non-perturbative corrections is rather small.
We expect that our simple implementation of non-perturbative corrections to fail in the peak (plateau) region, where one should retain more information about the shape function. Therefore, we only plot our NLL curves with non-perturbative corrections down to $t\sim 2\cdot 10^{-3}$ and $\tpa \sim 10^{-3}$, respectively. 

\begin{figure}
\centering
\includegraphics[width=0.49 \textwidth]{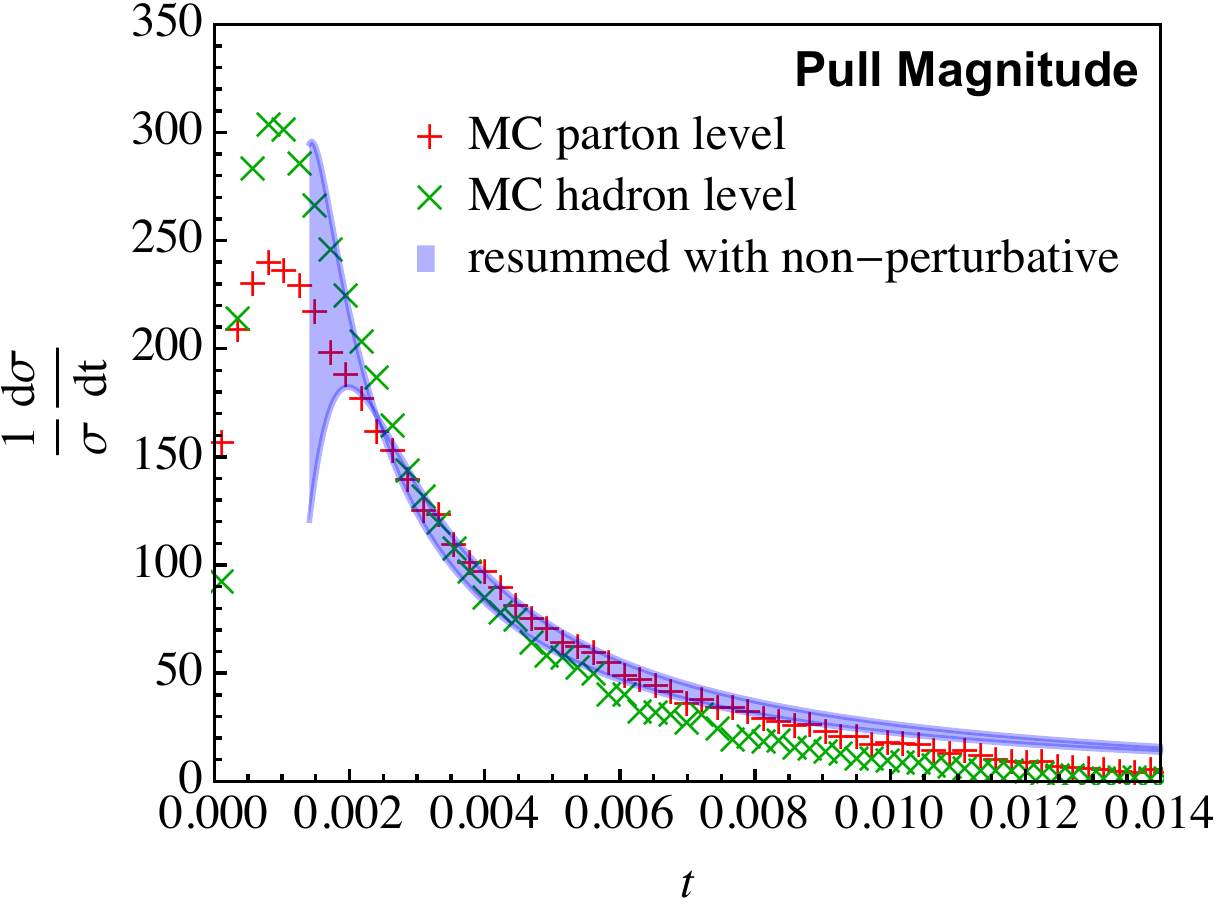}
\includegraphics[width=0.49 \textwidth]{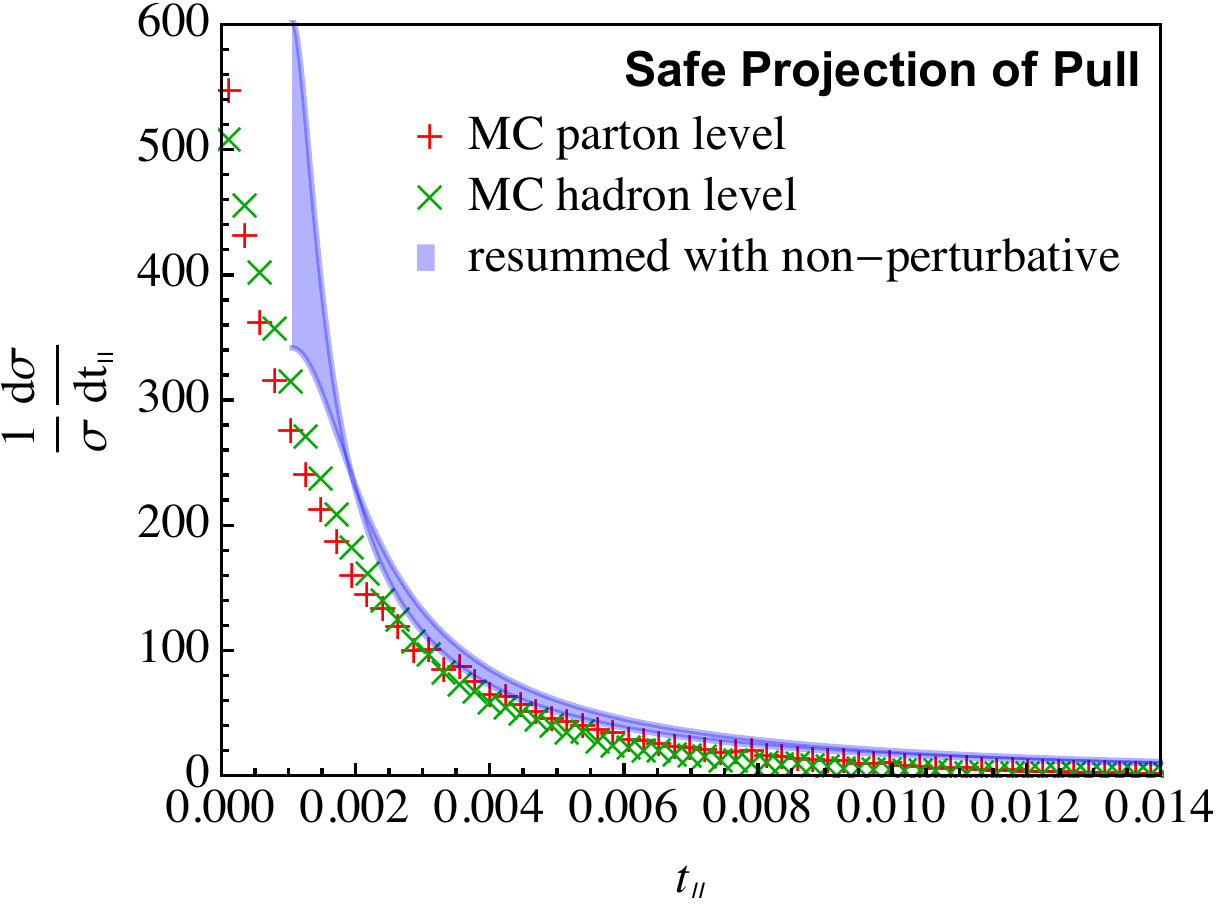}
\caption{Comparison of the distributions computed at NLL and supplemented with non-perturbative corrections, to a numerical simulation obtain with the event generator Pythia v8.240. 
}
\label{mc-comparisons} 
\end{figure}

In Fig.~\ref{mc-comparisons}, we compare our results to those obtained with a general-purpose Monte Carlo event generator. We generate a single event $pp\to H Z$ at $\sqrt{s}=13$~TeV, with the Higgs decaying in $b \bar b$ and $Z$ leptonically, using MadGraph v2.6.6~\cite{Alwall:2014hca} and we then shower this event many times in Pythia v8.240~\cite{Sjostrand:2014zea}. FastJet v3.3.2~\cite{Cacciari:2011ma} is used to find jets and calculate the pull variables. 
The Monte Carlo results for $t$ and $\tpa$ are then compared to our NLL predictions, supplemented by the non-perturbative corrections. 
We find decent agreement between the Monte Carlo and our NLL prediction for $t$ and $\tpa$, supplemented by non-perturbative corrections.
We note that the NLL and Monte Carlo predictions depart at the tail of the distributions. This effect is more noticeable for the pull magnitude and it signals the fact that the resummation alone is not enough to describe the distribution at large $t$ and matching to fixed-order is needed.

Finally, we expect additional non-perturbative contributions from the Underlying Event, due to multiple parton-parton interactions and pileup, due to multiple proton-proton interactions per bunch crossing. We have not included these effects in our studies, but we anticipate that their scaling with the jet radius will be the same as FSR, that we did calculate in this paper, albeit with a different, non-perturbative, coefficient.

\section{Asymmetries}\label{sec:asym}

The projections of the pull vector we have discussed thus far exhibit nice theoretical properties. In particular, as discussed at length, IRC safety ensures perturbative calculability, while non-perturbative contributions can be treated as (power) corrections. Furthermore, the particular definitions of the projections, see Eq.~(\ref{pull-proj}) resulted in observables that share many similarities in their all-order behaviour with variables that are among the most-studied in particle physics, such as the transverse momentum of a vector boson and its projections. However, we cannot fail to notice that presence of the absolute value in Eq.~(\ref{pull-proj}) leads to a loss of information. For instance, an emission in rapidity-azimuth region between the two jets and an emission outside, could potentially contribute to the same value of $\tpp$ or $\tpa$. 
Therefore, in order to fully exploit the radiation pattern, we can construct asymmetric distributions by directly considering the projections of the pull vector along the two directions of interest, i.e.\ $\vec{t}\cdot \hat{n}_\parallel$ and $\vec{t}\cdot \hat{n}_\perp$. We note that the dot products, as opposed to $\tpa$ and $\tpp$, are not positive-definite.

\begin{figure}
\centering
\includegraphics[width=0.49 \textwidth]{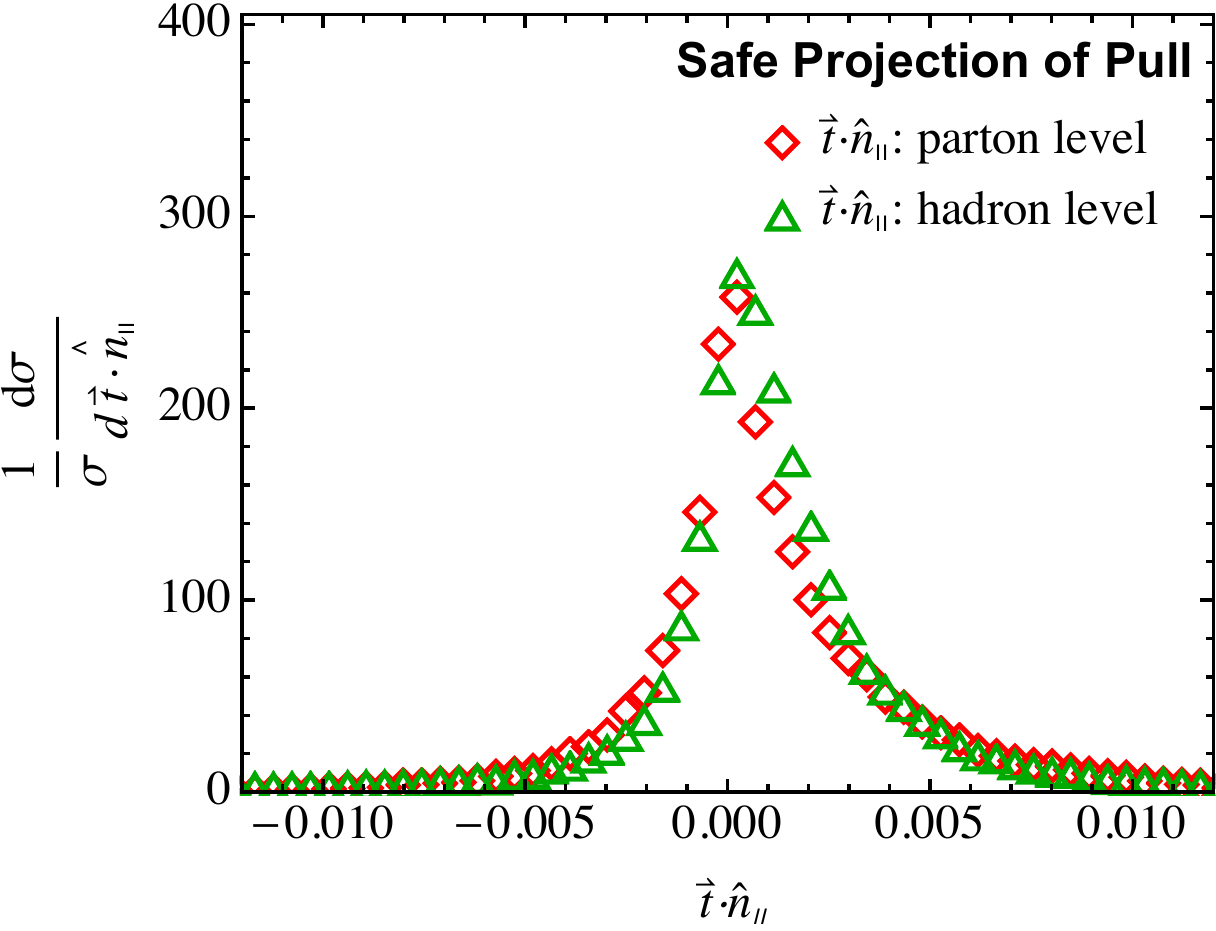}
\includegraphics[width=0.49 \textwidth]{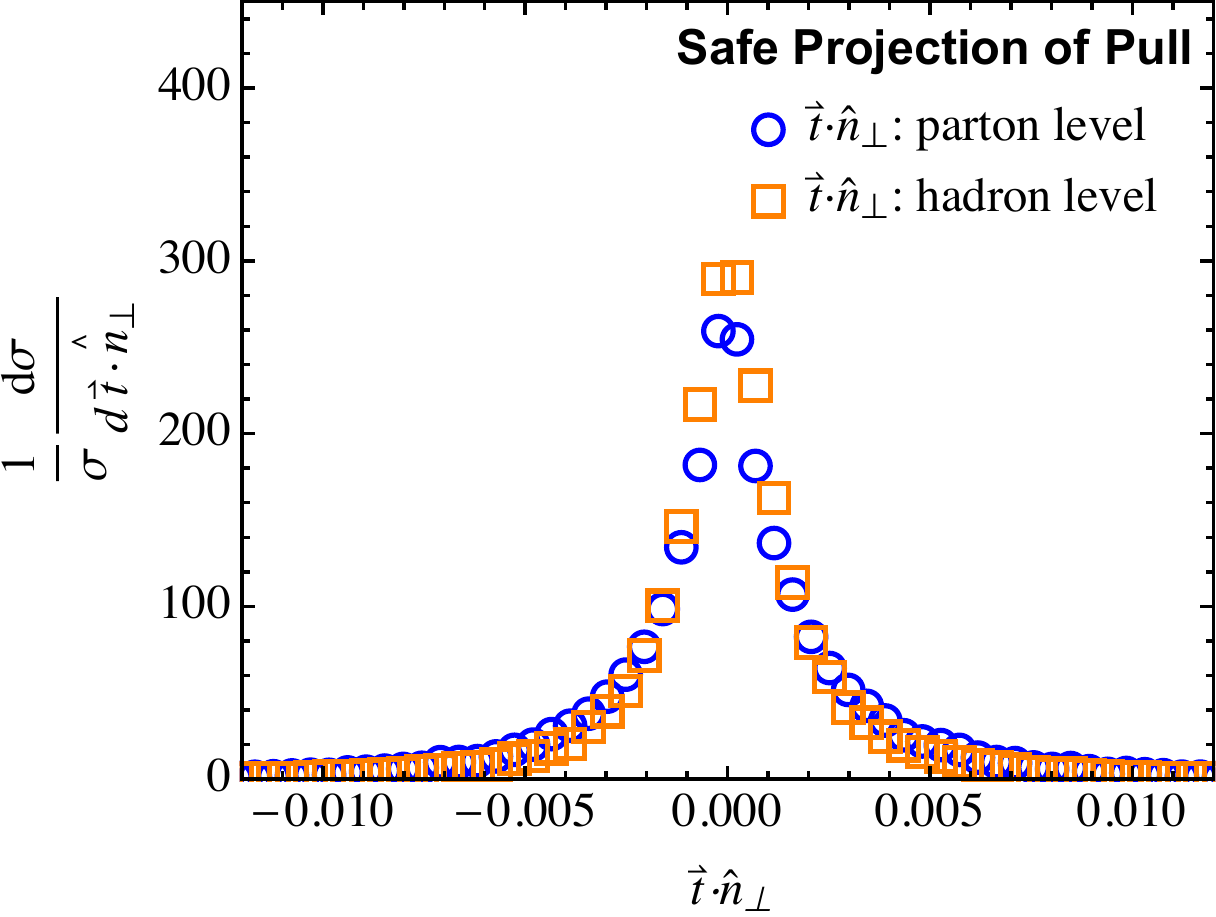}
\caption{Monte-Carlo simulations of the $\vec{t}\cdot \hat{n}_\parallel$ and $\vec{t}\cdot \hat{n}_\perp$ distributions, left and right respectively, measured on $H \to b \bar b$ events generated with Pythia~v8.240. The plots show results at both parton and hadron level.} 
\label{mc-asymmetric-distributions} 
\end{figure}

\begin{figure}
\centering
\includegraphics[width=0.48 \textwidth]{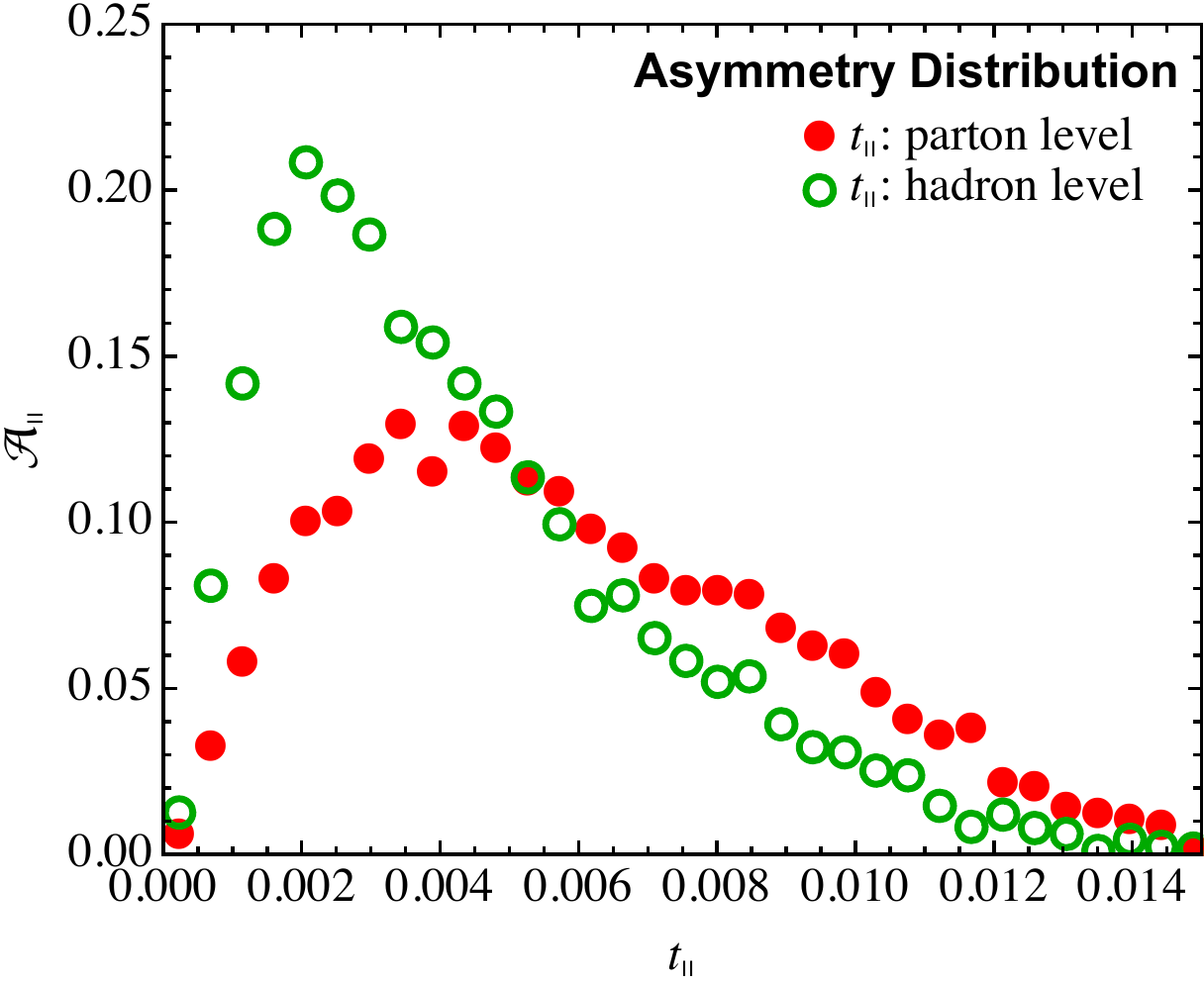}
\includegraphics[width=0.49 \textwidth]{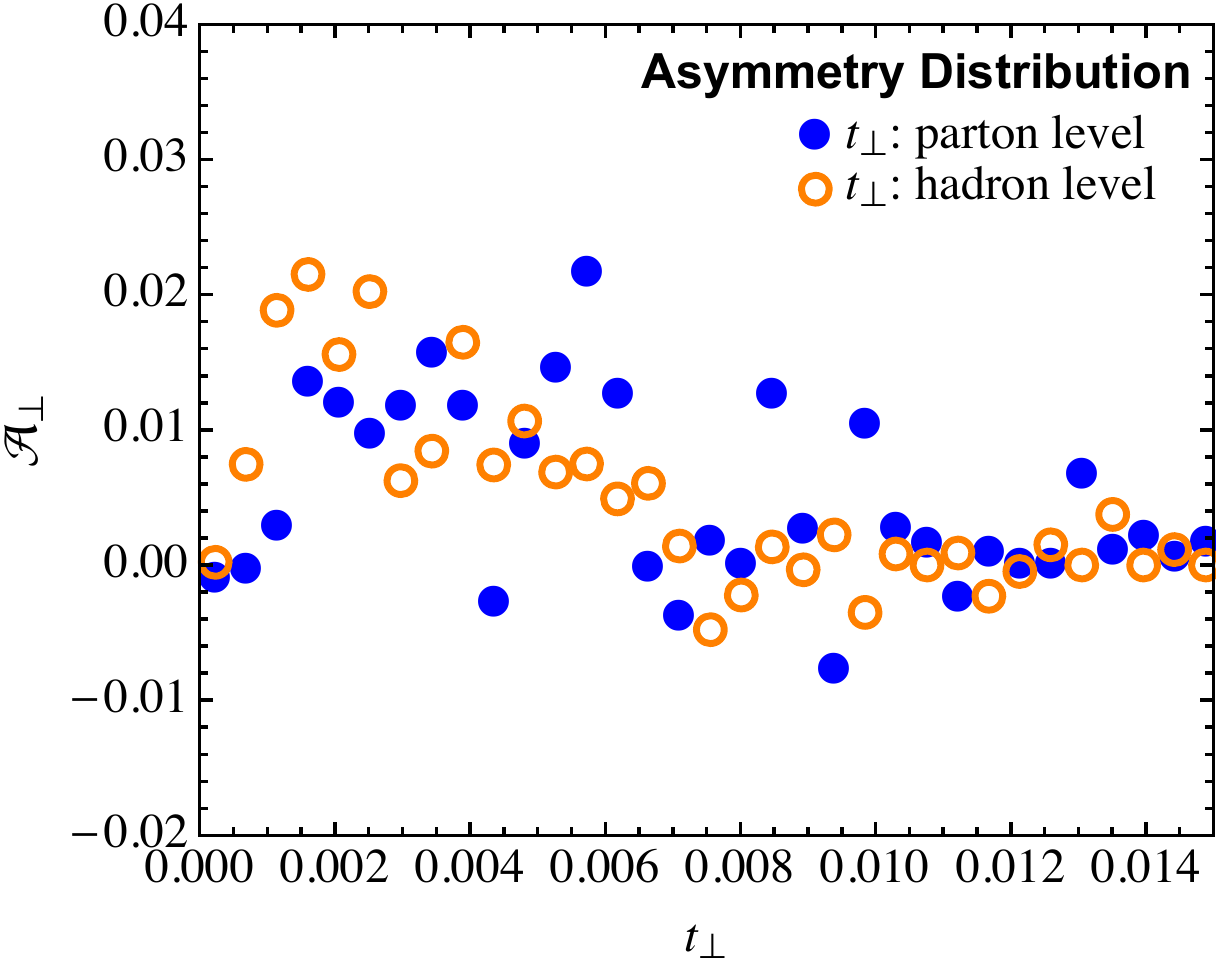}
\caption{Monte-Carlo simulations of the $\mathcal{A}_\parallel$ and $\mathcal{A}_\perp$ distributions, left and right respectively, measured on $H \to b \bar b$ events generated with Pythia~v8.240. The plots show results at both parton and hadron level.} 
\label{mc-asymmetries} 
\end{figure}

In Fig.~\ref{mc-asymmetric-distributions} we perform a Monte-Carlo study of these distributions for the colour singlet decay $H \to b \bar b $, using again the event generator Pythia~v8.240, with the same kinematical settings of the previous section. For each distribution we show both parton-level and hadron-level results. 
We would expect the $\vec{t}\cdot \hat{n}_\perp$ to be roughly symmetric about zero, while the distribution of $\vec{t}\cdot \hat{n}_\parallel$ should be skewed in the direction of the colour-connected leg of the dipole, here the positive direction. The plots show that this is indeed the case. 
In order to emphasise these features even more, we can build the following asymmetry distributions
\begin{align}\label{asym_def}
\mathcal{A}_\parallel&=\frac{t_\parallel}{\sigma}\frac{d \sigma}{d t_\parallel}\Big|_{\vec{t}\cdot \hat{n}_\parallel>0}-\quad \frac{t_\parallel}{\sigma}\frac{d \sigma}{d t_\parallel}\Big|_{\vec{t}\cdot \hat{n}_\parallel<0}, \\
\mathcal{A}_\perp&=\frac{t_\perp}{\sigma}\frac{d \sigma}{d t_\perp}\Big|_{\vec{t}\cdot \hat{n}_\perp>0}-\quad \frac{t_\perp}{\sigma}\frac{d \sigma}{d t_\perp}\Big|_{\vec{t}\cdot \hat{n}_\perp<0}
\end{align}
We expect $\mathcal{A}_\parallel$ to be more marked than $\mathcal{A}_\perp$ and this is indeed what is found in the simulations, as shown in Fig.~\ref{mc-asymmetries}.

We note that the above asymmetries are still IRC safe and therefore can be calculated in perturbation theory. 
Indeed, we could argue that $\mathcal{A}_\parallel$ is essentially the IRC safe version of the pull angle distribution. 
The definitions of the asymmetries in Eq.~(\ref{asym_def}) make explicit references to the sign of the scalar product which is used to project the pull vector. This constraint essentially introduces a new boundary in phase-space which renders the all-order structure of these observables richer. While we expect that this resummation can still be achieved, in this work we limit ourselves to analytically evaluate the asymmetries at fixed-order.
The lowest-order contribution to the asymmetries originates from wide-angle soft emissions. In particular, we find that the contribution denoted by $\mathcal{A}$ in Eq.~(\ref{dipole-expanded}) does not vanish when we integrated separately over the $\vec{t}\cdot \hat{n}_i>0$ and $\vec{t}\cdot \hat{n}_i<0$ regions. 
We find
\begin{align}
\mathcal{A}_\parallel&=\frac{\as C_F}{ \pi}\left[ \frac{4R}{\pi} \frac{\cos \beta \sinh \Delta y+ \sin \beta \sin \Delta \phi}{\cos \Delta \phi-\cosh \Delta y} + \mathcal{O} \left( R^3\right) \right]
 +\mathcal{O} \left( \as^2\right)
, \\
\mathcal{A}_\perp&= \frac{\as C_F}{ \pi} \left[ \frac{4R}{\pi} \frac{\cos\beta \sin \Delta \phi-\sin \beta\sinh \Delta y}{\cos \Delta \phi-\cosh \Delta y}+ \mathcal{O} \left( R^3\right) \right]
 +\mathcal{O} \left( \as^2\right).
\end{align}
Interestingly, the asymmetries are sensitive to odd powers of the jet radius, in the small-$R$ expansion. This comes about because of the restrictions on the angular integrations imposed by the $\vec{t}\cdot \hat{n}_i>0$ and $\vec{t}\cdot \hat{n}_i<0$ constraints. 
We also point out that these asymmetries essentially depend on soft radiation, while collinear contributions cancel out. Soft radiation exhibit strong sensitivity to the pattern of colour correlations and therefore these observables can provide a valuable testing ground for Monte Carlo parton showers that attempt to go beyond the large-$N_c$ limit, e.g.~\cite{Nagy:2019pjp,Forshaw:2019ver}.

\section{Conclusions and Outlook}\label{sec:conclusions}
A detailed understanding of colour flow in hard scattering processes is of primary interest for LHC phenomenology for numerous reasons. First of all, it provides a valuable way of separating hadronic decay products of colour singlets, such as the Higgs or any other electroweak bosons, from the QCD background, often originating from gluon splittings. 
Furthermore, should new strongly-interacting states be discovered at the LHC, colour correlations can be used to characterise the colour representation these particles live in.
However, precision studies of colour flow in hadron-hadron collisions are challenging because of the sensitivity to the soft and non-perturbative regimes of QCD. Therefore, it is important to devise observables that, while maintaining the desired sensitivity, offer theoretical robustness. 
In this context, infra-red and collinear safety is an important requirement because it ensures perturbative calculability, with dependence on non-perturbative corrections that is, at least parametrically, under control.  Perturbative calculations for IRC safe observables can be used, in turn, to test the ability of general-purpose Monte Carlo event generators to correctly simulate colour flow in proton-proton collisions at hight energy. 

In this study we have considered the observable jet pull, which has been introduced in order to probe colour flow between hard jets.
Measurements of the pull angle have been advocated as sensitive probe of inter-jet radiation and have been performed at the Tevatron and the LHC. 
In particular, precision measurements by the ATLAS collaboration challenges the ability of general-purpose Monte Carlo event generators to correctly describe these distributions. In a previous Letter, we addressed the theoretical calculation of the pull angle distribution but we found difficult to draw firm theoretical conclusions due fact that the pull angle is not an IRC safe observable. 

In this current paper, we have put forward novel observables that aim to probe colour flow in an efficient way, while featuring IRC safety. In particular, we have noticed that while the pull angle, i.e.\ the angle between the pull vector and the line joining the centres of the jets of interest, in the azimuth-rapidity plane, is not IRC safe, the projections of the pull vector along ($\tpa$) and orthogonal to ($\tpp$) such an axis are. 
Therefore, these observables can be computed in perturbation theory. We have performed all-order calculations for these two projections and, for comparison, for the magnitude of the pull vector, considering the interesting case of a Higgs boson decaying into a pair of bottom quarks.
Our results are valid to next-to-leading logarithmic accuracy, in the limit where the considered observable is small. 
In this context, besides collinear radiation, we have also investigated the structure of soft-emissions at wide angle and of non-global logarithms, expressing our results as a power series in the jet radius. 
Matching to fixed-order perturbation theory is possible but we have left it for future work. Furthermore, we have supplemented our results with an estimate of non-perturbative corrections arising from the hadronisation process and compared our results to simulations obtained with a Monte Carlo parton shower.

The theoretical understanding reached in this study has led us to the introduction of novel asymmetry distributions that measure the radiation pattern by looking at the difference between the jet pull vector pointing towards and away from the other jet of interest. 
In particular, the asymmetry distribution $\mathcal{A}_\parallel$ can be considered the IRC version of the pull angle distribution. We have pointed out that such asymmetries can have interesting applications both in the context of tagging colour singlets, such as $H\to b\bar b$ versus $g \to b \bar b$, and as a means to test how general-purpose Monte Carlo event generators probe soft emissions beyond the leading colour approximation. Therefore, we look forward to study these asymmetries in more detail in order to arrive to their all-order resummation.

Finally, we note that it would be interesting to study observables sensitive to colour flow in the rest frame of the decaying particle, in order to minimise kinematical effects originating from asymmetric decays. The standard definition of jet pull does not seem appropriate for this kind of studies, however the modification we put forward in Ref.~\cite{Larkoski:2019urm} appears to be better suited. We look forward to continue our investigation in this direction too.

\acknowledgments
We thank Andy Buckley and Giovanni Ridolfi for useful discussions, Ben Nachman, Matthew Schwartz, and Jesse Thaler for comments on the manuscript. 
This work is partly supported by the curiosity-driven grant ``Using jets to challenge the Standard Model of particle physics" from Universit\`a di Genova. 

\bibliography{pull}

\end{document}